\documentclass[twocolumn]{aastex6}

\usepackage{graphicx}
\usepackage{array}
\usepackage{amssymb}
\usepackage{natbib}
\usepackage{float}
\usepackage{color}
\usepackage[utf8]{inputenc}  
\usepackage{amsmath}  
\usepackage{comment}
\usepackage{footnote}
\usepackage{longtable}
\usepackage[caption2]{ccaption}
\usepackage[figuresright]{rotating}
\usepackage{booktabs}

\renewcommand*{\thefootnote}{\fnsymbol{footnote}}

\shorttitle{Varaibility of narrow and broad line Seyfert 1 galaxies}
\shortauthors{Suvendu Rakshit and C. S. Stalin}

\begin{document}


\title{Optical variability of narrow and broad line Seyfert 1 galaxies}



\author{Suvendu Rakshit and C. S. Stalin}
\affil{Indian Institute of Astrophysics, Block II, Koramangala, Bangalore-560034, India}
\email{Email: suvenduat@gmail.com}

\begin{abstract}
We studied optical variability (OV) of a large sample of 
narrow-line Seyfert 1 (NLSy1) and broad-line Seyfert 1 (BLSy1) galaxies 
with $z$ $<$ 0.8 to investigate any differences in their OV properties. 
Using archival optical $V$-band light curves from the Catalina Real 
Time Transient Survey that span $5-9$ years and modeling them 
using damped random walk, we 
estimated the amplitude of variability. We found NLSy1 galaxies
as a class show lower amplitude of variability than their broad-line 
counterparts. In the sample of both NLSy1 and BLSy1 galaxies,  radio-loud 
sources are found to have higher variability amplitude than radio-quiet 
sources. Considering only sources that are detected in the X-ray band, 
NLSy1 galaxies are less optically variable than BLSy1 galaxies. The amplitude of 
variability in the sample of both NLSy1 and BLSy1 galaxies is found to be anti-correlated with Fe II strength but correlated 
with the width of the H$\beta$ line. The well-known anti-correlation 
of variability-luminosity and the variability-Eddington ratio is present in our data. 
Among the radio-loud sample, variability amplitude is found to be correlated 
with radio-loudness and radio-power suggesting jets also play an 
important role in the OV in radio-loud objects, in addition to the 
Eddington ratio, which is the main driving 
factor of OV in radio-quiet sources. 
\end{abstract}
\keywords{galaxies: active - galaxies: Seyfert - techniques: photometric}




\section{Introduction}
Active galactic nuclei (AGNs) are among the luminous extragalactic sources
in the sky persistently emitting radiation with bolometric luminosities 
as large as 10$^{48}$ erg s$^{-1}$ \citep{2002ApJ...579..530W}. They are believed
to be powered by accretion of matter onto supermassive black holes
(SMBH) at the center of galaxies \citep{1969Natur.223..690L,1984ARA&A..22..471R}. A small fraction ($\sim$15\%) 
of AGNs are radio-loud and emit in the radio band, thereby possessing powerful 
relativistic jets \citep{1989AJ.....98.1195K}. One of the important observed
characteristics of all categories of AGN is that they show variations in 
their emitted flux.
This was known since their discovery as a class of object \citep{1963Natur.197.1040S,1963Natur.197.1041G}. The flux variations in AGN are random and occur on different 
time scales of minutes, hours and days, and has been observed over the 
complete accessible wavelengths \citep{1995ARA&A..33..163W,1997ARA&A..35..445U}. 

In the recent years, AGNs have been extensively studied for their optical variability (OV). Several theoretical models have
been proposed to explain the observed flux variations such as 
accretion disk instabilities \citep{1998ApJ...504..671K}, multiple 
supernovae \citep{1997MNRAS.286..271A}, micro-lensing 
\citep{2000A&AS..143..465H}, Poisson process 
\citep{2000ApJ...544..123C} and damped random walk  \citep[DRW;][]{2009ApJ...698..895K}. However, we still do not have an understanding
of the underlying physical processes that cause flux
variability. Extensive optical photometric monitoring of large samples of AGN has revealed important connections between the observed
variability and the various important physical 
properties of the sources. Some of the observed correlations are 
the dependency of the amplitude of variability with wavelength 
\citep{1996ApJ...463..466D}, luminosity 
\citep{1994MNRAS.268..305H,2009ApJ...698..895K,2011A&A...525A..37M}, redshift \citep{2004ApJ...601..692V,2011A&A...525A..37M}, black hole mass and Eddington ratio \citep{2007MNRAS.375..989W,2009ApJ...696.1241B,2009ApJ...698..895K,2010ApJ...721.1014M}. 
Variability is thus an important tool to investigate the complex nature
of the central engine and accretion processes in AGN.

Past variability studies mainly focused on broad line AGN, however, 
only limited reports are available in the literature on the OV characteristics
of Narrow Line Seyfert 1 (NLSy1) galaxies.  NLSy1 galaxies are a peculiar type 
of Seyfert 1 galaxies having full width at half maximum (FWHM) of the permitted line FWHM$\mathrm{(H\beta) 
<2000\, km\,s^{-1}}$ and flux ratio of [O III] to H$\beta <3$ 
\citep{1985ApJ...297..166O,1989ApJ...342..224G}. They usually show strong Fe II 
emission compared to their broad line counterparts \citep{2001A&A...372..730V}. 
NLSy1 galaxies also show strong soft X-ray variability and steep X-ray spectra 
than  
BLSy1 galaxies \citep{1999ApJS..125..317L,2004AJ....127.1799G}.  
They harbor low mass black hole (10$^6$ - 10$^8\, M_{\odot}$) and have high Eddington ratio compared
to the BLSy1 galaxies that are believed to be hosted by heavier ($\gtrsim$ 10$^8 \, M_{\odot}$) black holes \citep{2006ApJS..166..128Z,2012AJ....143...83X}. However, recent 
studies indicate that  NLSy1 galaxies do have black hole masses similar to 
that of blazars \citep{2013MNRAS.431..210C,2016MNRAS.458L..69B}. Several authors have studied 
the reasons for mass deficit in NLSy1 galaxies and suggest 
geometrical factors to be the reason for observing narrow emission lines
and consequently leading to low black hole mass determination from virial estimates
\citep{2008MNRAS.386L..15D,2013MNRAS.431..210C,2016IJAA....6..166L}. Though 
there are some differences in the  observed properties of BLSy1 and NLSy1 
galaxies, we do
not as of now have a clear picture on the similarities and/or differences in the OV properties
between these two classes of objects. A comparative analysis of the OV
properties of BLSy1 and NLSy1 galaxies could provide clues to the cause of the 
peculiar observational characteristics of NLSy1 galaxies.   

Though NLSy1 galaxies have not been studied extensively for 
optical variability, a few studies do exist in the literature. Such studies
though limited to a handful of sources have focused both on 
flux variations within a night \citep{1999MNRAS.304L..46Y,2000NewAR..44..539M} 
as well as on time scales of days 
\citep{1999MNRAS.304L..46Y,2000NewAR..44..539M}.   
\citet{2006ARep...50..708D} performed long-term optical photometric monitoring 
of a NLSy1 galaxy Ark 564 and found a low variability amplitude of $0.1-0.2$ 
mag. Similar results have also been found by \citet{2004ApJ...609...69K} who 
studied 6 NLSy1 galaxies and concluded that NLSy1 galaxies as a class show 
less variability than BLSy1 galaxies and the extreme variability seen in the 
soft X-ray is not present in the optical. Both these studies lack a proper 
sample of NLSy1 galaxies. Benefited from the Solan Digital Sky Survey (SDSS) 
multi-wavelength, multi-epoch-repeated photometric data and the extended 
catalog of NLSy1 galaxies compiled by \citet{2006ApJS..166..128Z}, a 
comparative study of NLSy1 and BLSy1  galaxies with a moderate sample of 55 
NLSy1 galaxies and a control sample of 108 BLSy1 galaxies was first made by 
\citet{2010ApJ...716L..31A} and subsequently by \citet{2013AJ....145...90A}.
They found: (1) NLSy1 galaxies have systematically smaller variability than 
BLSy1 galaxies; (2) 
the amplitude of variability increases with the width of 
H$\beta$ and strength of [O III] lines but decreases with the strength of 
Fe II emission; (3) variability is anti-correlated with Eddington ratio 
but insignificant with luminosity; (4) a positive correlation with black hole mass is found which, however, vanishes after controlling Eddington ratio in the analysis, which 
the authors noted could be due to the limited ranges of luminosity and black 
hole mass of their sample. However, these findings are based on the poorly sampled SDSS photometric light curves having only about 27 observations over a duration of 5 years.

Earlier studies on the OV of a large sample
of NLSy1 galaxies were limited because (i) a small number of NLSy1 galaxies known 
at that time and (ii) the lack of long-term photometric data. Recently, \citet{2017ApJS..229...39R} have compiled a new catalog of NLSy1 galaxies consisting of 11,101 sources, which is a factor of five increase in the 
number of NLSy1 galaxies than the previous catalog. During the course of NLSy1 
galaxies selection, \citet{2017ApJS..229...39R} have also arrived at a large sample 
of BLSy1 galaxies. Long-term $V$-band observations of many of these objects are available from the Catalina Real Time Transient 
Survey \citep[CRTS;][]{2009ApJ...696..870D}. Motivated by the availability of large sample of BLSy1 and 
NLSy1 galaxies and the CRTS data, we carried out a comparative study of OV of both BLSy1 and NLSy1 galaxies to understand 
the correlation of variability amplitude to different physical characteristics of these 
two populations of sources. In this paper, we present the results of
this study. This paper is organized as follows. In section \ref{sec:data}, we 
present the sample of NLSy1 and BLSy1 galaxies and the photometric data used 
for this study followed by the analysis presented in section \ref{sec:analysis}. The results are given in section \ref{sec:result} followed by a summary and conclusion in 
section \ref{sec:conclusion}. A cosmology with $H_0 = 70\, \mathrm{km\,s^{-1}\, Mpc^{-1}}$, $\Omega_\mathrm{m} = 0.3$, and $\Omega_{\lambda} = 0.7$ is assumed throughout.

\section{Sample and data}\label{sec:data}
\subsection{NLSy1 and BLSy1 galaxies sample}
Our sample of NLSy1 galaxies is taken from \citet{2017ApJS..229...39R}. This was extracted from a systematic re-analysis of the spectra of objects in SDSS DR12 \citep{2015ApJS..219...12A} that are classified as ``QSO'' by the automatic SDSS spectroscopic pipeline \citep{2002AJ....123.2945R}.
The custom made emission line fitting process by \citet{2017ApJS..229...39R}
to identify new NLSy1 galaxies from SDSS DR12 
database also allowed them to compile a sample of BLSy1 galaxies that have FWHM
(H$\beta$) $>2200 \,\mathrm{km\, s^{-1}}$. Since the number of BLSy1 galaxies 
is very large, for this work, we created a sub-sample 
of BLSy1 galaxies having median SNR $>10 \, \mathrm{pixel^{-1}}$ in the SDSS spectra resulting in 14,894 BLSy1 
galaxies. Thus, the sample of sources selected for this study consists of 11,101 
NLSy1 galaxies and 14,894 BLSy1 galaxies. 

\subsection{Photometric data}
 
For OV studies of the sample selected above, we used
the data from CRTS\footnote{\url{http://nesssi.cacr.caltech.edu/DataRelease}} \citep{2009ApJ...696..870D}. It provides light curves with much higher temporal sampling thereby enabling us to study the variability characteristic of our sample 
\citep{2015MNRAS.453.1562G}. CRTS streams data from three telescopes; the 
0.7 m Catalina Schmidt Telescope with a field of view of 8 deg$^2$, the 
1.5 m Mount Lemmon Survey reflector telescope having a 1 deg$^2$ field of 
view located at north of Tucson, Arizona, and the 0.5 m Uppsala
Schmidt telescope at Siding Spring, Australia having a 4.2 deg$^2$ field of
view. CRTS covers about 2500 deg$^2$ sky per night taking 4 exposures per 
visit that are separated by 10 min. Observations are made over 21 nights per 
lunation reaching $V$-band magnitude of around $19-20$ mag. All data are processed in 
real time using an automated software and calibrated to Johnson $V$-band. 
CRTS data contains light 
curves of about 500 million sources. The detailed information regarding CRTS 
survey and the optical light curves can be found in \citet{2009ApJ...696..870D,2013ApJ...763...32D} and \citet{2015MNRAS.453.1562G}. 

We cross-correlated our sample of sources with CRTS within a search 
radius of 3$\arcsec$. This cross-correlation yielded optical light curves
for a reduced sample of 9069 NLSy1 galaxies and 13,928 BLSy1 galaxies. A common 
practice in dealing with light curves obtained in large surveys is to 
identify and consequently remove any spurious outliers that might have 
been caused by photometric or technical errors. To remove such outliers we applied an iterative 3$\sigma$ clipping algorithm around local group of data points to all the light curves. For any given light curve we removed points with more than 3$\sigma$ deviation 
from the mean and repeat this process until no points with 3$\sigma$ deviation 
is present in the light curve or the number of points in the light curve 
between two consecutive iterations remain the same. Since our aim is to study OV, we further considered only those light curves that have
a minimum of 50 epochs of data. This brings down the sample size to 
9063 NLSy1 and 13,831 BLSy1 galaxies.

\begin{figure}
\centering
\resizebox{8.5cm}{7.0cm}{\includegraphics{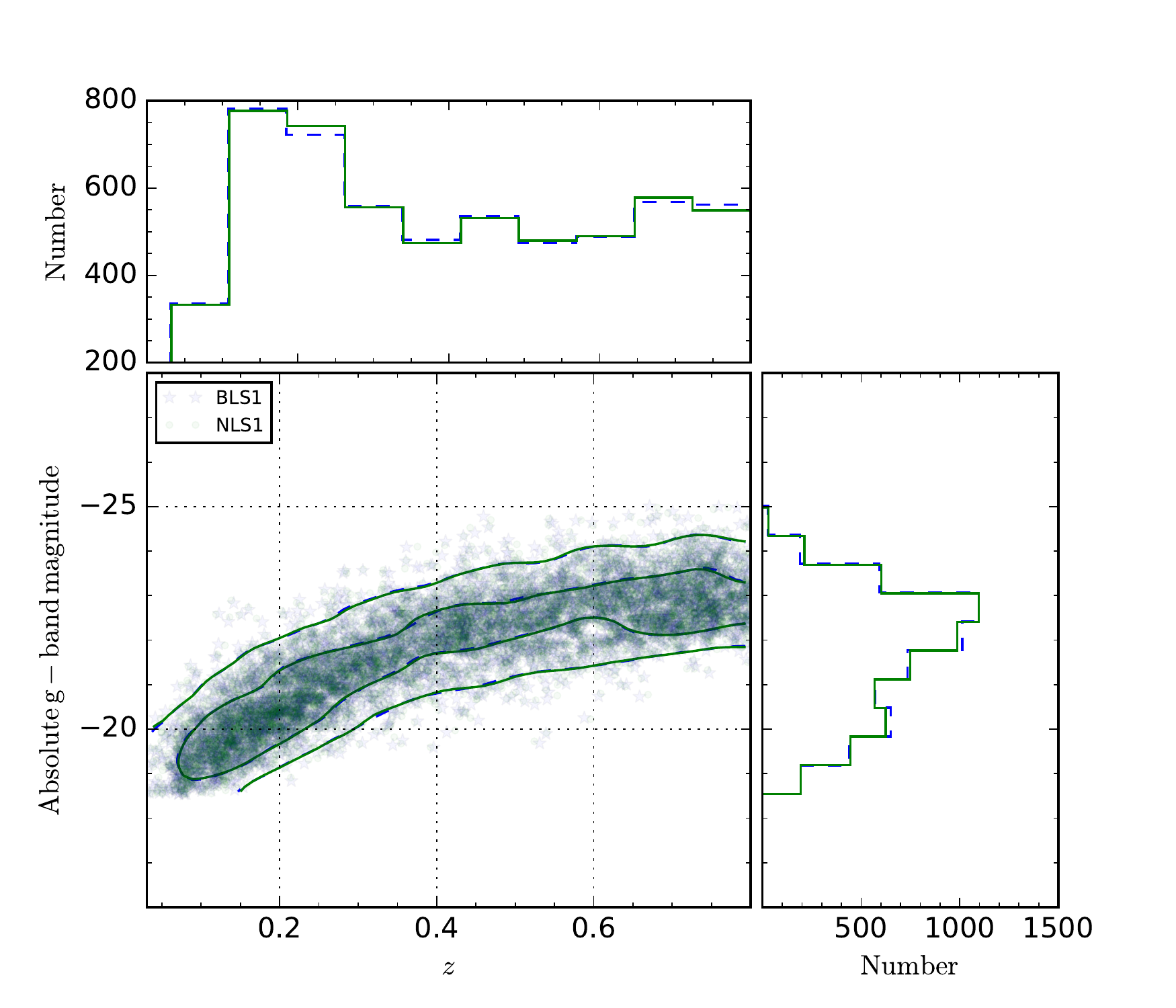}}
\caption{Absolute $g$-band magnitude ($M_g$) against redshift ($z$) for BLSy1 galaxies ( dashed contours) and NLSy1 galaxies (solid contours). The contours are the 68 and 95 percentile density contours. The corresponding distribution of $z$ (top) and $M_g$ (right) is also shown. Both NLSy1 galaxies (solid line) and BLSy1 galaxies (dashed line) have similar distributions.}\label{Fig:Mg_z}. 
\end{figure} 

\begin{figure*}
\centering
\resizebox{5.5cm}{3cm}{\includegraphics{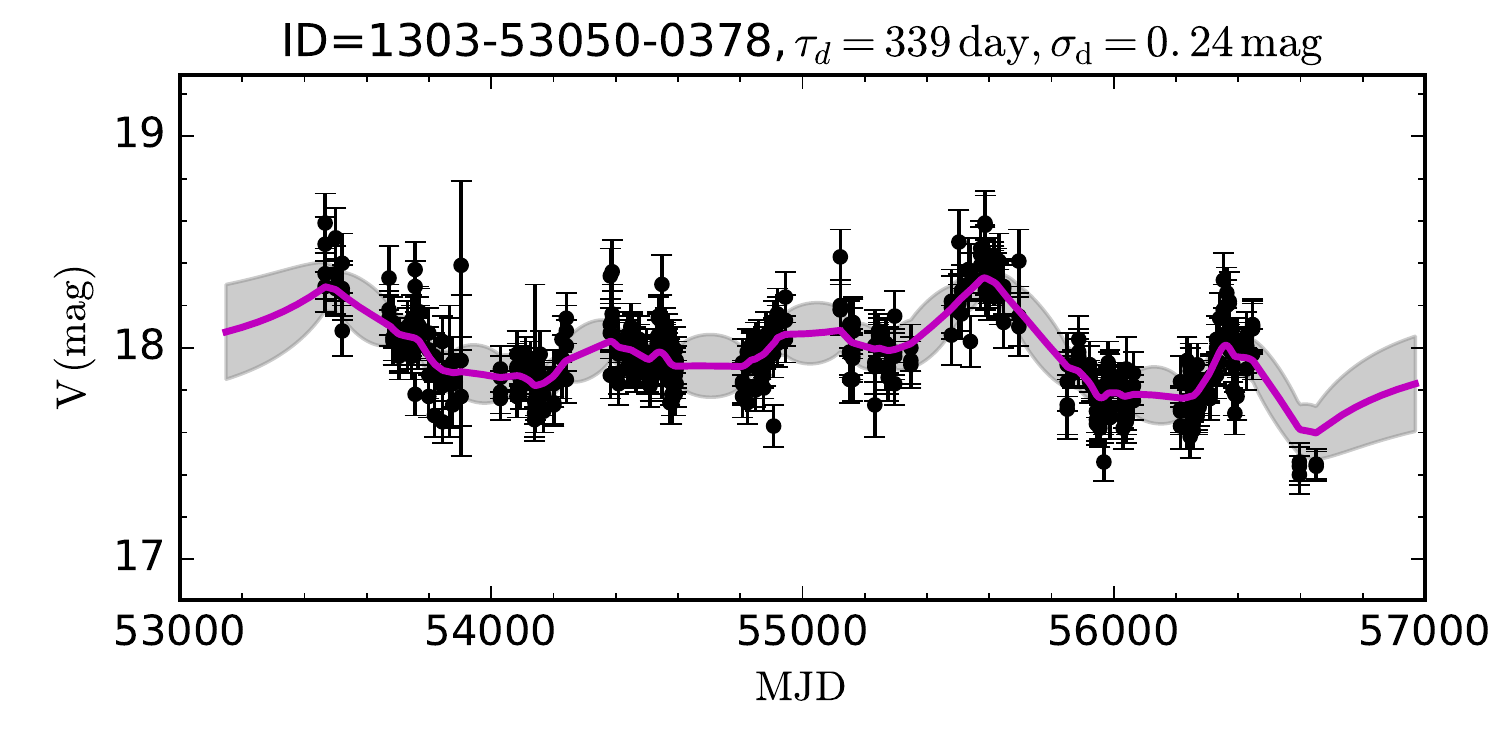}}
\resizebox{5.5cm}{3cm}{\includegraphics{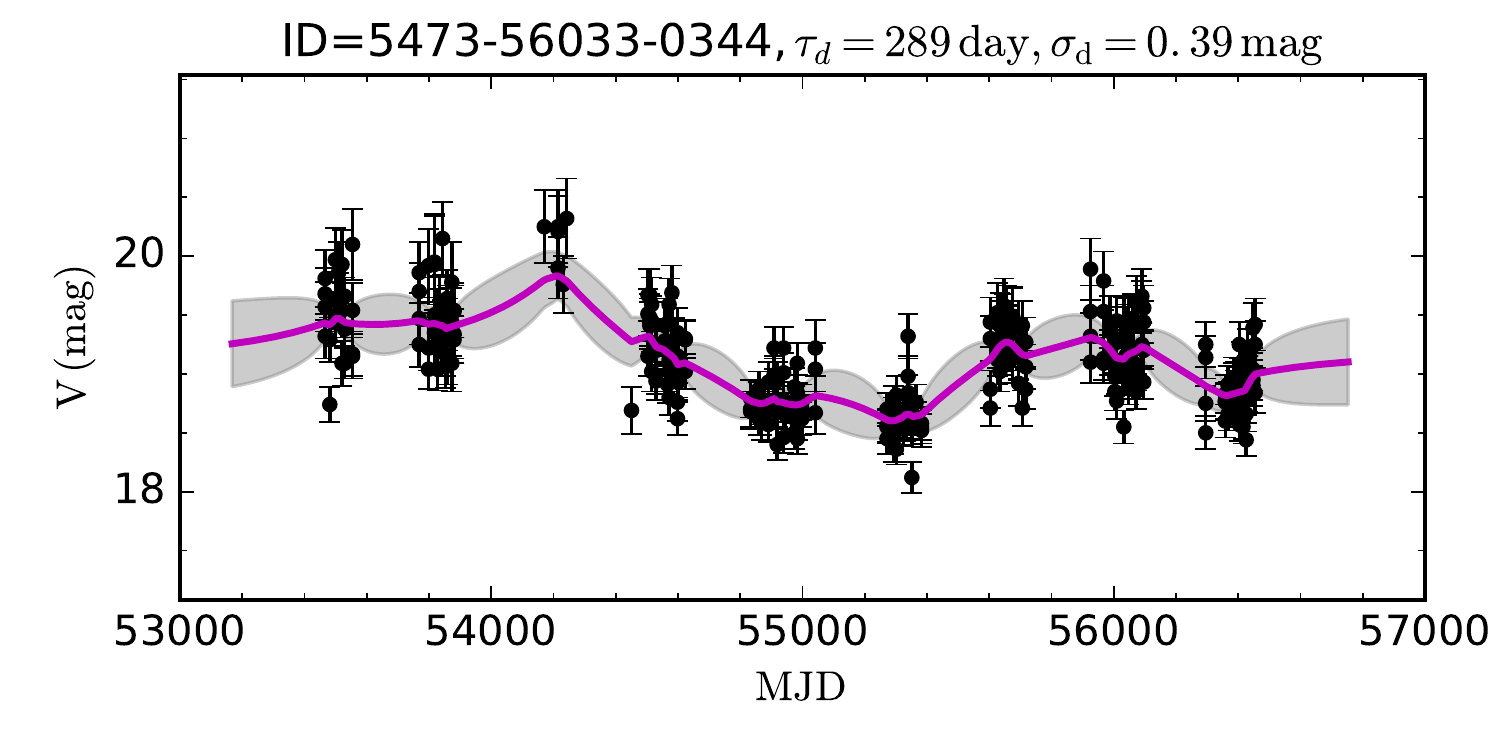}}
\resizebox{5.5cm}{3cm}{\includegraphics{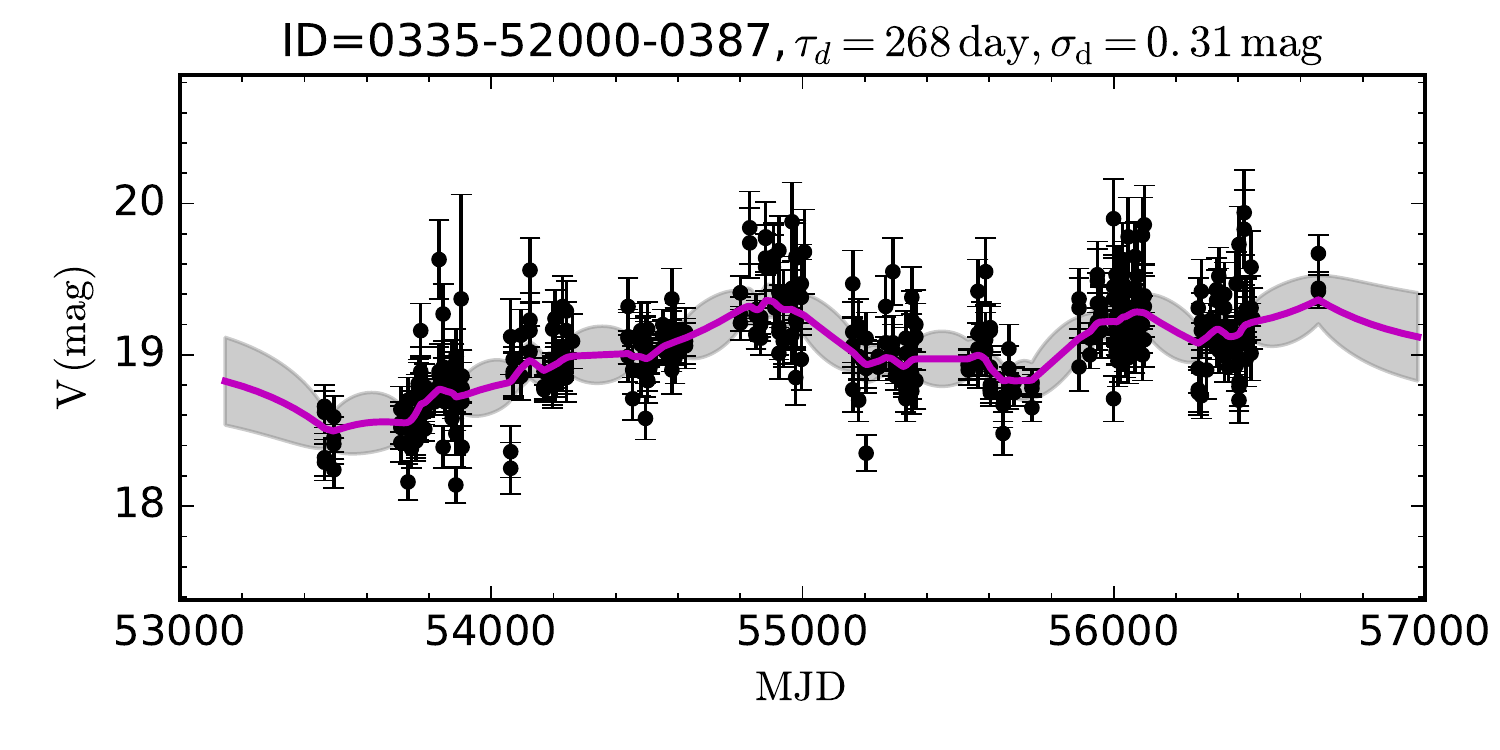}}
\resizebox{5.5cm}{3cm}{\includegraphics{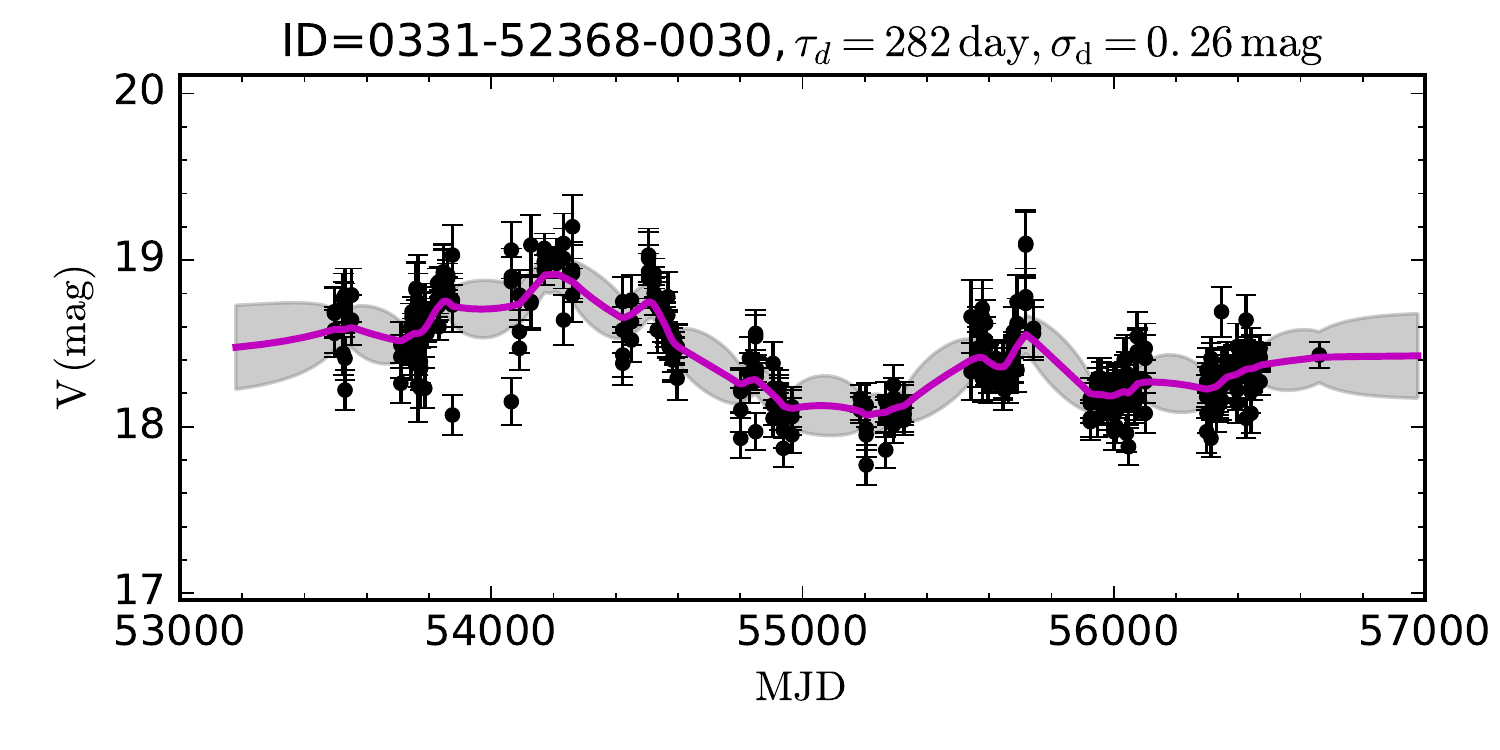}}
\resizebox{5.5cm}{3cm}{\includegraphics{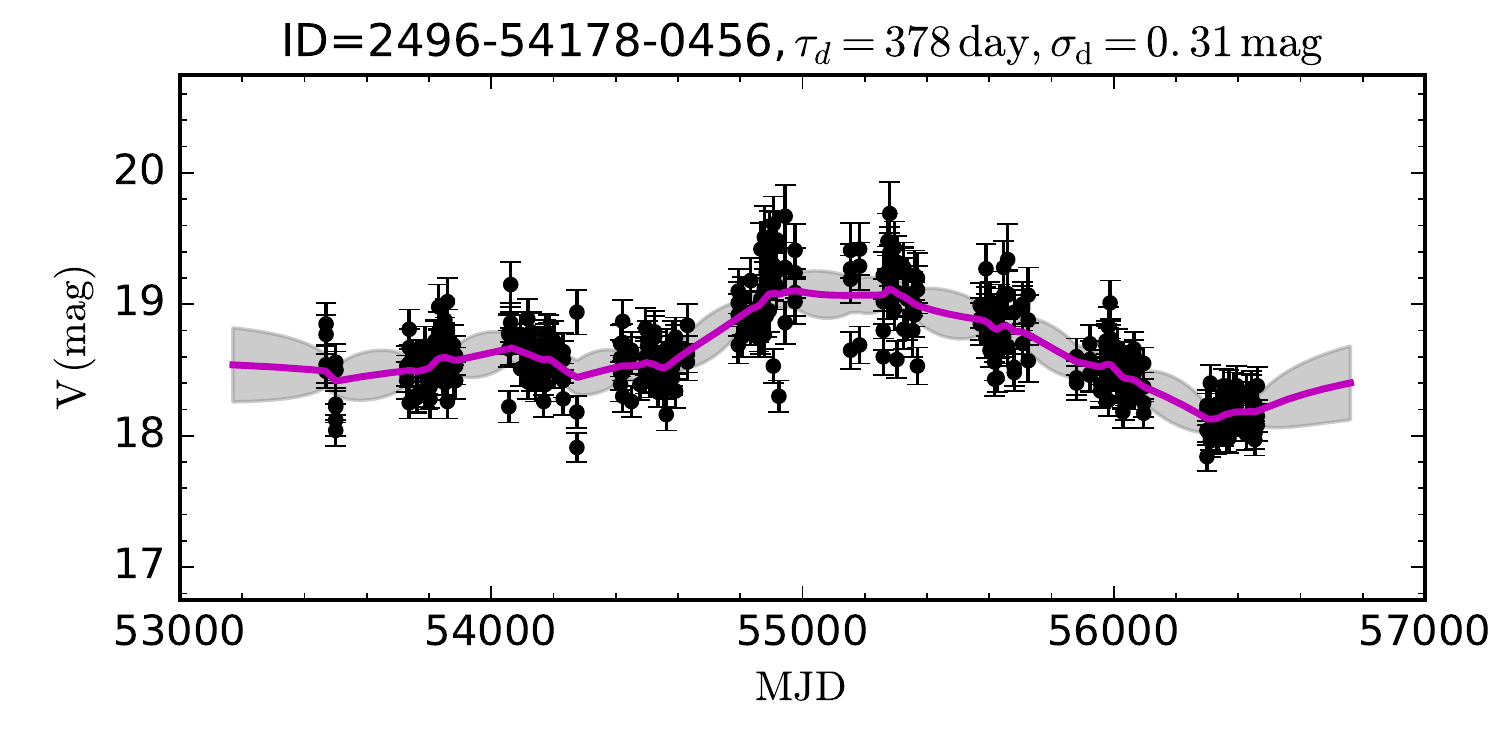}}
\resizebox{5.5cm}{3cm}{\includegraphics{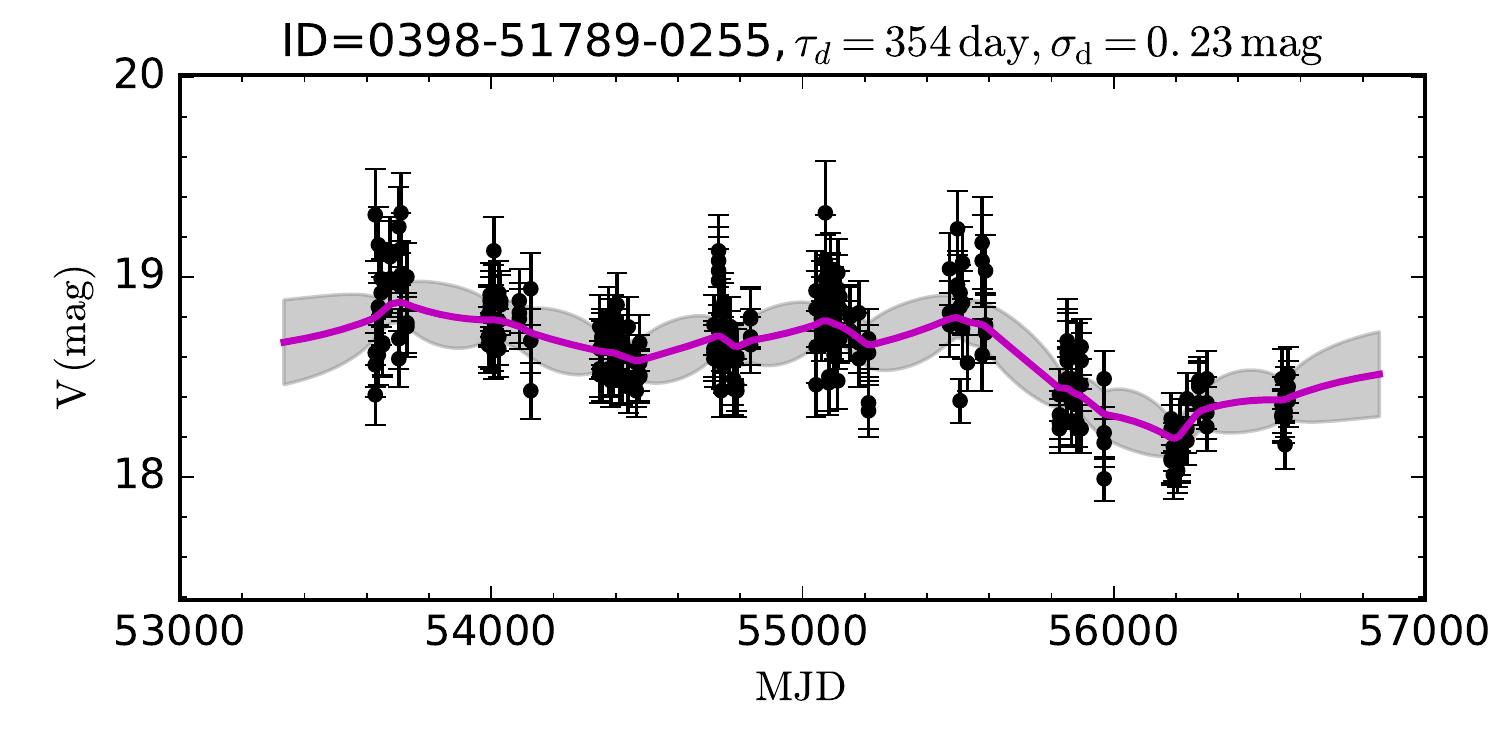}}
\caption{A few examples of light curve fitting using JAVELIN. The black points with error bar are the CRTS data and the solid line shows the best fit of the light curve while the shaded area is the 1$\sigma$ error region. The SDSS ID (plate-MJD-fiber) and the JAVELIN best-fitted parameters (observed frame) are noted in each panel.}\label{Fig:javelin_fit}. 
\end{figure*}

The motivation of this work is to carry out a systematic comparative study of the OV properties between NLSy1 and BLSy1 galaxies. It is, therefore, 
imperative that the sample so selected must match as close as possible to each other in the  luminosity-redshift plane.  For that, we divided both the samples in small redshift and luminosity bins, and randomly selected an equal number of NLSy1 and BLSy1 galaxies from each bin. This resulted in a final 
working sample of 5510 NLSy1 and  5510 BLSy1 galaxies. The distribution of the sample of NLSy1 and BLSy1 galaxies ($z<0.8$) in the luminosity-redshift plane is shown in Figure \ref{Fig:Mg_z}. From the figure, it is evident that both the sample of galaxies resemble  each other. The two-dimensional  
Kolmogorov–Smirnov (K-S) test \citep{1992nrfa.book.....P} yields a statistics of 0.007 and a $p$-value of 0.98 confirming that both samples have the same $M_g-z$ distribution.

\section{Analysis of variability}\label{sec:analysis}
The CRTS $V$-band\footnote{The $V$-band has an effective wavelength ($\lambda_{\mathrm{eff}}$) of 5510 \AA\, with a bandwidth of 880 \AA\, thus, covering a wavelength range of about 4600 \AA\, to 6400 \AA. It can, thus, be contaminated by strong broad emission lines such as redshifted Mg II ($\lambda_{\mathrm{eff}} =2800$ \AA) and H$\beta$ ($\lambda_{\mathrm{eff}}= 4861$ \AA), which also vary with time and follow the nuclear continuum variations. Since our sample spans $z=0$ to 0.8, H$\beta$ will contribute to the $V$-band flux for the objects having $z<0.31$ while Mg II will contribute for objects having $z>0.65$. Therefore, it is very difficult to disentangle the relative contribution of broad emission lines and the continuum to the broad $V$-band photometry.} light curves of the sources studied for variability contain
a minimum of 50 epochs of data and the total duration of the light curves
span $5-9$ years.
Various studies show that optical variability of quasar can be well described by a damped random walk 
\citep[DRW;][]{2009ApJ...698..895K}, which is a stochastic process with an exponential covariance function 
\begin{equation}
S(\Delta t)=\sigma_{d}^{2} \exp \left(-\frac{|\Delta t|}{\tau_d} \right),
\end{equation}
where $\sigma_d$ is the amplitude and $\tau_d$ is a characteristic time scale 
of variability \citep{2010ApJ...708..927K,2010ApJ...721.1014M,2011ApJ...735...80Z,2013ApJ...765..106Z,2016ApJ...819..122Z}. DRW is shown 
to provide a realistic explanation of quasar optical variability and both the 
model parameters, $\sigma_d$ and $\tau_d$, are correlated with the physical 
parameters of AGN such as luminosity, black hole mass and Eddington ratio 
\citep{2009ApJ...698..895K,2010ApJ...708..927K,2010ApJ...721.1014M,2013A&A...554A.137A,2016ApJ...826..118K}. 
DRW model is a powerful tool to quantify variability characteristics on time scales of several days to years although recently some limitations of the DRW model has been noticed especially when dealing with light curves 
that have duration of observations about ten times shorter than the true DRW timescale \citep{2016MNRAS.459.2787K,2016arXiv161108248K}.
            
\begin{figure}
\centering
\resizebox{8cm}{7.0cm}{\includegraphics{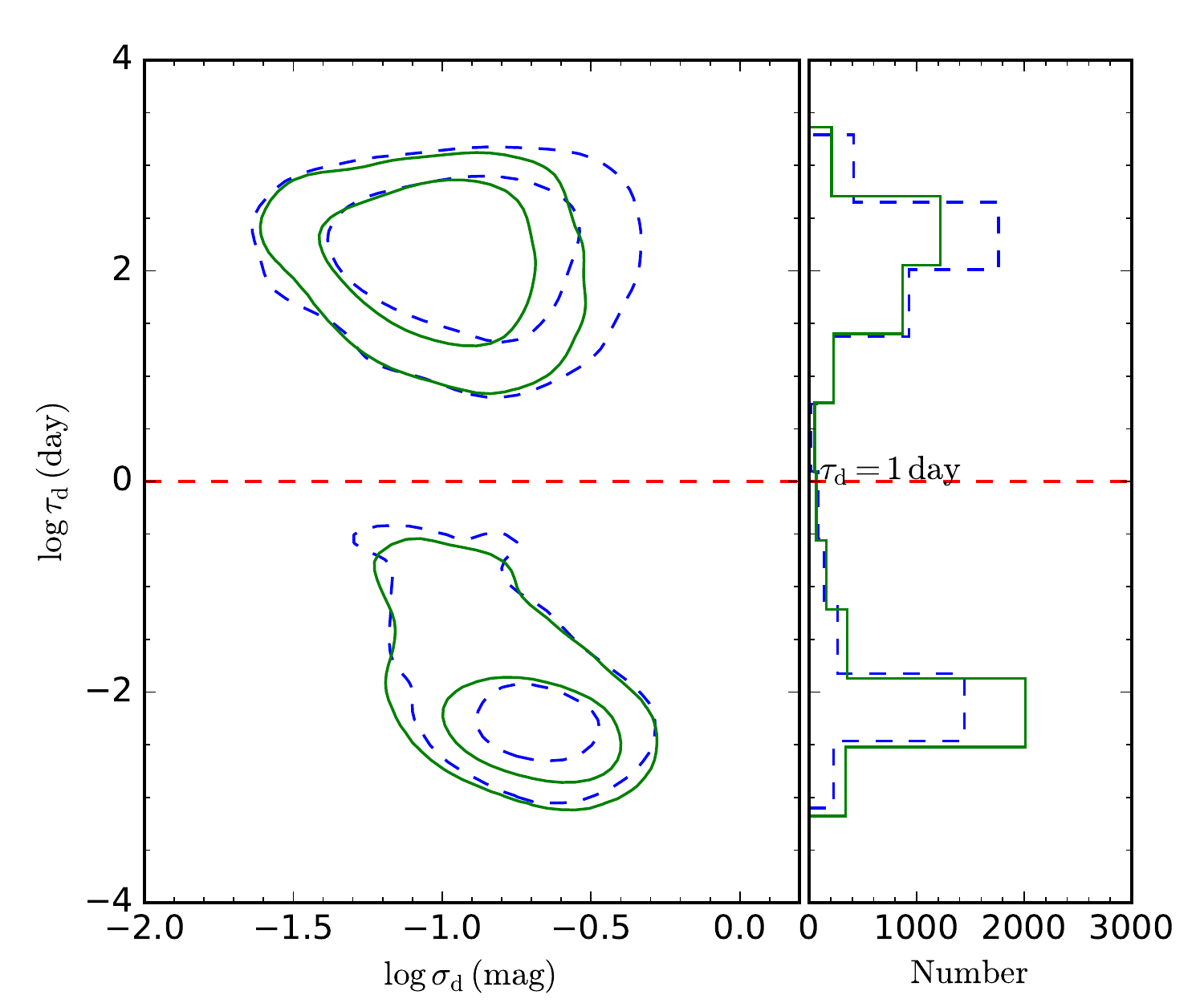}}
\caption{The distribution of JAVELIN best-fitted parameters, the rest frame damping time scale ($\tau_d$) and the observed frame amplitude of variation $\sigma_d$, as obtained from the fitting of NLSy1 galaxies (solid contours) and BLSy1 galaxies (dashed contours) in the $\tau_d-\sigma_d$ plane. The shown contours are the 68 and 95 percentile density contours. The horizontal-dashed line indicates $\tau_d=1$. The distribution of $\tau_d$ (1D cut along the y-axis) is shown on the left panel for NLSy1 (solid line) and BLSy1 (dashed line) galaxies.}\label{Fig:javelin_params}. 
\end{figure}

To quantify the variability characteristics of all the sources in our 
sample, we fit each of their $V$-band CRTS light curves using DRW model 
implemented in JAVELIN\footnote{\url{http://bitbucket.org/nye17/javelin}} which is 
a python code developed by \citet{2011ApJ...735...80Z}. Logarithmic priors for both $\tau_d$ and $\sigma_d$ have been used for fitting the light curves. JAVELIN has been used widely in literature to model the continuum and line 
emission light curves of AGN reverberation mapping data \citep{2012ApJ...744L...4G,2014MNRAS.445.3073P}. Few examples of fitting done
on the light curves are shown in Figure \ref{Fig:javelin_fit}. The estimated 
values of the parameters after fitting all light curves is plotted in 
Figure \ref{Fig:javelin_params}, where $\tau_d$ is the rest frame time scale 
i.e., the time scale of variability corrected for the redshift. A bimodal 
distribution along the time axis is clearly visible and delineated by a gap at about 1 day. Since the time sampling of CRTS is larger than 1 day, any variability 
on time scales shorter than 1 day is unreliable and cannot be used for variability analysis. Also, these sources have poor quality light curves with no noticeable variability trend as confirmed by visual examinations. Therefore, only objects having $\tau_d$ greater than 1 day was considered for
further variability analysis. This leads us to a sample of  2161 (39.2\%) 
NLSy1 and 2919 (52.9\%) BLSy1 galaxies for further variability analysis. This 
also suggests that only 39\% of NLSy1 galaxies from our original sample 
are variable while 61\% NLSy1 galaxies are non-variable on time scales
larger than a day. However, in the case of BLSy1 galaxies, about 53\% of our original sample is variable 
which implies that overall BLSy1 galaxies are more variable than NLSy1 
galaxies on time scales longer than a day.

Following \citet{2010ApJ...716L..31A}, we also calculated the intrinsic 
amplitude of variability ($\sigma_m$). This was estimated from
the measured variance of the observed light curves  after subtracting the 
measurement errors. The $\sigma_m$ is estimated using the following 
formalism \citep[see][]{2007AJ....134.2236S}
\begin{equation}
\Sigma=\sqrt{\frac{1}{n-1}\sum_{i=1}^{N}(m_i - <m>)^2},
\end{equation}
where $<m>$ is the weighted average and	the amplitude of variability  
$\sigma_m$ is             
\[\sigma_m  =
  \begin{cases}
    \sqrt{\Sigma^2 - \epsilon^2},  & \quad \text{if } \Sigma>\epsilon,\\
     0,                            & \quad  \text{otherwise.}\\
  \end{cases}
\]

\begin{figure}
\centering
\resizebox{7cm}{6.0cm}{\includegraphics{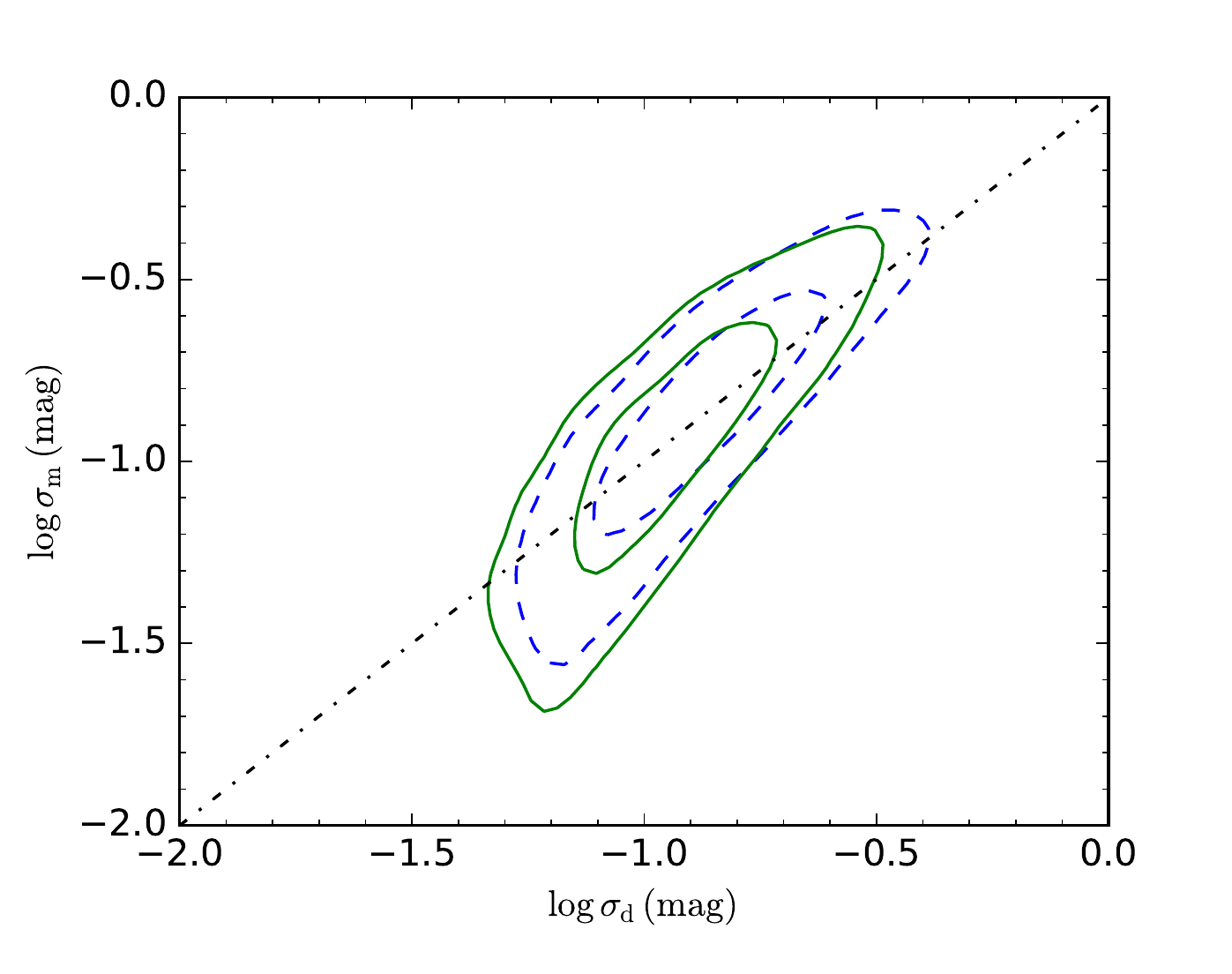}}
\caption{Comparison between two variability amplitude indicators, $\sigma_m$ and $\sigma_d$ in the observed frame for NLSy1 galaxies (solid contours) and BLSy1 galaxies (dashed contours). The shown contours are the 68 and 95 percentile density contours. The dashed-dot line represents one to one correspondence between $\sigma_m$ and $\sigma_d$.}\label{Fig:sigd_psi}. 
\end{figure}

Here, $\epsilon$ represents the contribution of measurement errors to the 
variance and it is estimated directly from the errors of individual observed 
magnitudes $\epsilon_i$,
\begin{equation}
\epsilon^2=\frac{1}{N}\sum_{i=i}^{N} \epsilon_{i}^{2}. 
\end{equation}
In Figure \ref{Fig:sigd_psi}, we show the two indicators  of variability 
amplitude for our sample of sources, $\sigma_d$ obtained from fitting 
the light curves using JAVELIN and $\sigma_m$ estimated directly from the 
observed light curves. The dashed-dot line indicates one to one correspondence between the two values. 
In this work, we used $\sigma_d$ as the indicator of the amplitude of variation unless specified otherwise.

\begin{figure}
\centering
\resizebox{8.5cm}{7.5cm}{\includegraphics{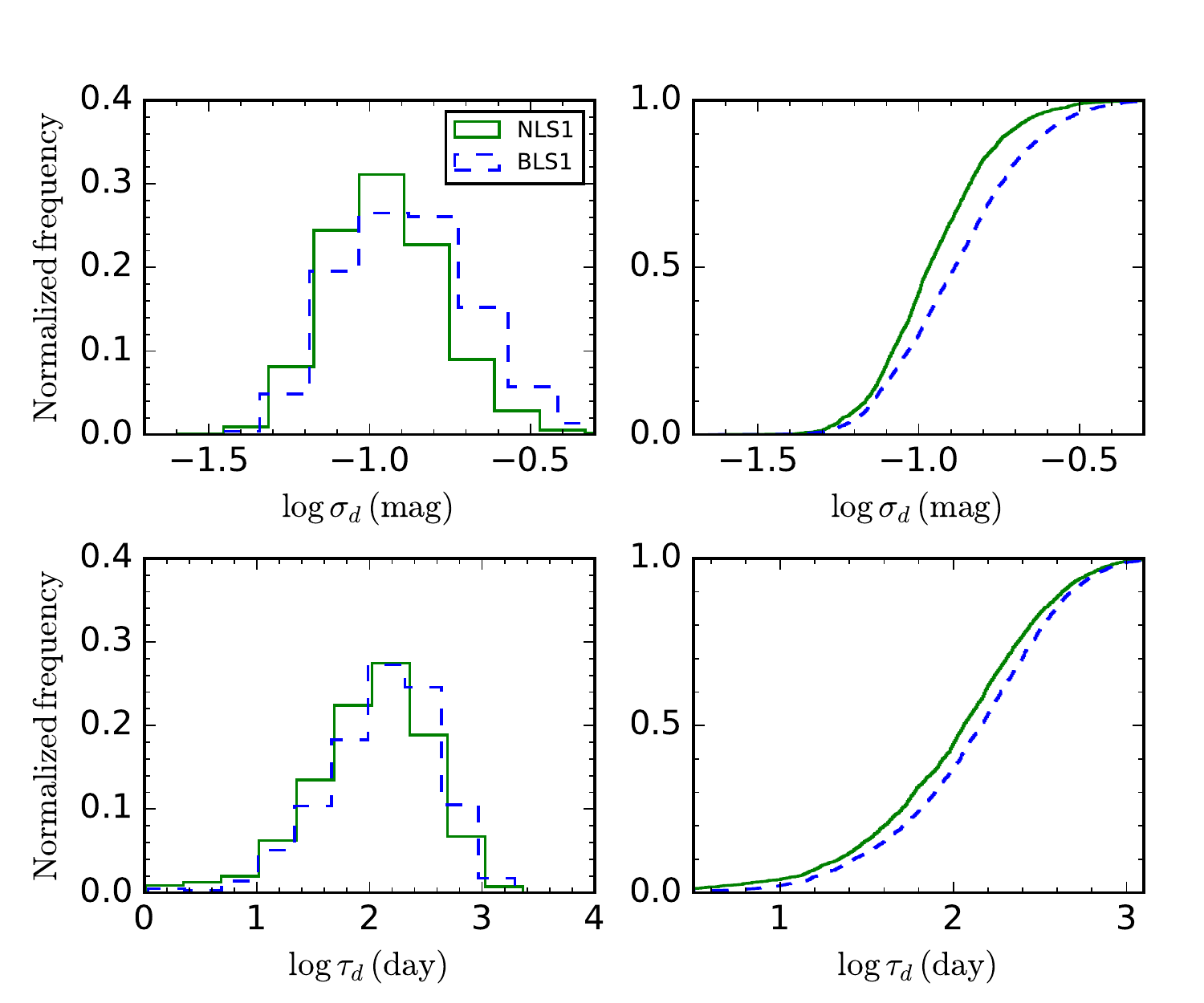}}
\caption{Upper panel: The distribution of variability amplitude ($\sigma_d$; left panel) and its cumulative distribution (right panel). Lower panel: The same for the characteristic time scale of variability. The solid line is for NLSy1 galaxies and dashed line is for BLSy1 galaxies.}\label{Fig:BLS1_vs_NLS1}. 
\end{figure}

\section{Results}\label{sec:result}
\subsection{NLSy1 vs BLSy1 galaxies}

We compare the variability amplitudes and time scales for the 2161 NLSy1 and 2919 BLSy1 galaxies in Figure \ref{Fig:BLS1_vs_NLS1}. The normalized histogram (left panel) and cumulative distribution (right panel) of variability amplitude are shown in the upper panels while the same for time scales are shown in the lower panels.  
The $\sigma_d$ distribution has a median of $0.107^{+0.057}_{-0.032}$ mag and $0.129^{+0.082}_{-0.049}$ mag for NLSy1 and BLSy1 galaxies respectively. When $\sigma_m$ distribution is considered, we find the median values are $0.112^{+0.089}_{-0.056}$ and $0.136^{+0.114}_{-0.069}$ for NLSy1 and BLSy1 galaxies respectively. Within errors, the median amplitude of variability found in BLSy1 and
NLSy1 are similar, however, a two-sample K-S test applied to the 
distribution of  $\sigma_d$ for NLSy1 and BLSy1 
galaxies yields a $D$-statistic value  of 0.17 and $p$-value of 6 $\times$
10$^{-31}$,
confirming that the two distributions are significantly different. NLSy1 galaxies as a class are thus less variable than BLSy1 galaxies. Therefore, the strong variability shown by NLSy1 galaxies in X-rays 
relative to BLSy1 galaxies is not seen in the optical band. 
The distribution of time scales for NLSy1 galaxies has a peak slightly smaller 
than that of BLSy1 galaxies, having a median value of $116^{+207}_{-83}$ days. This is smaller than the median value of $146^{+235}_{-104}$ days found in BLSy1 galaxies, though the 
scatter in the distribution is very large. A two-sample K-S test of $\tau_d$ distribution for NLSy1 and BLSy1 galaxies yields a $D$-statistic value  of 0.09 and $p$-value of 2 $\times$
10$^{-8}$. As the $D$-statistics is small it is difficult to draw any conclusion on the distribution of $\tau_d$ between NLSy1 and BLSy1 galaxies. According to \citet{2016arXiv161108248K} for reliable estimation of $\tau_d$, one needs to have data with a minimum duration of about 10 times longer than the true DRW time scale. As our data span 5 to 9 years, about 18\% of NLSy1 galaxies and 24\% of BLSy1 galaxies have the duration of light curve $<10\times \tau_d$. Thus, for those light curves $\tau_d$ may not represent the true time scale. However, $\sigma_d$ is independent of the duration of light curves and only affected by the photometric noise. As our main motivation is to have a comparative analysis of the amplitude of variability ($\sigma_d$) of NLSy1 and BLSy1 galaxies, we restricted ourselves to further analysis of $\sigma_d$ only.

This study using a matched sample of NLSy1 and BLSy1 galaxies has clearly
demonstrated the difference in their OV properties with NLSy1 galaxies
showing lower variability amplitude than BLSy1 galaxies. This finding is consistent with the results of
\citet{2004ApJ...609...69K} who studied optical variability using a small
sample and the ensemble variability study of
\citet{2010ApJ...716L..31A,2013AJ....145...90A} who analyzed a sample of 55
NLSy1 galaxies and a control sample of 108 BLSy1 galaxies. The weaker
variability of NLSy1 galaxies compared to BLSy1 galaxies can be understood in terms of them having smaller width and stronger Fe II emission compared to
BLSy1 galaxies.
As accretion
disk is normally thought to be responsible for optical/UV radiation from
AGN, the difference in the OV properties leads us to speculate differences in the physical processes operating in the accretion disks of NLSy1 and BLSy1 galaxies. One possibility could be the slim disk scenario in NLSy1 galaxies compared to the standard Shakara-Sunyaev, geometrically thin optically thick accretion disk in BLSy1 galaxies \citep{2013AJ....145...90A}.

\subsection{Radio subsample}\label{sec:radio}

The observed optical emission from radio-loud and radio-quiet objects can be due to a combination of the different physical process. One of the ways to ascertain this is to see if there is any difference in the OV properties of both NLSy1 and BLSy1 galaxies when they are subdivided based on their radio properties. We, therefore, cross-correlated our sample of NLSy1 and BLSy1 
galaxies with the Faint Images of the Radio Sky at Twenty centimeters 
(FIRST)\footnote{\url{http://sundog.stsci.edu}} catalog \citep{1995ApJ...450..559B}
within a search radius of 2$\arcsec$.
We found 122 out of 2161 NLSy1 galaxies and 276 out of 2919 BLSy1 galaxies
are detected in FIRST. Depending on their radio-loudness (defined as the 
logarithmic flux ratio of 1.4 GHz to $g$-band flux, i.e., 
$R=f_{1.4 \,\mathrm{GHz}}/f_g$), we further divided the radio detected NLSy1 and BLSy1 galaxies into radio-quiet (RQ; $\log R<1$) and radio-loud (RL; $\log R>1$) sub-categories. This 
resulted in 48 (54) radio-quiet and 74 (222) radio-loud NLSy1 (BLSy1) 
galaxies. 

\begin{figure}
\centering
\resizebox{8.5cm}{3.5cm}{\includegraphics{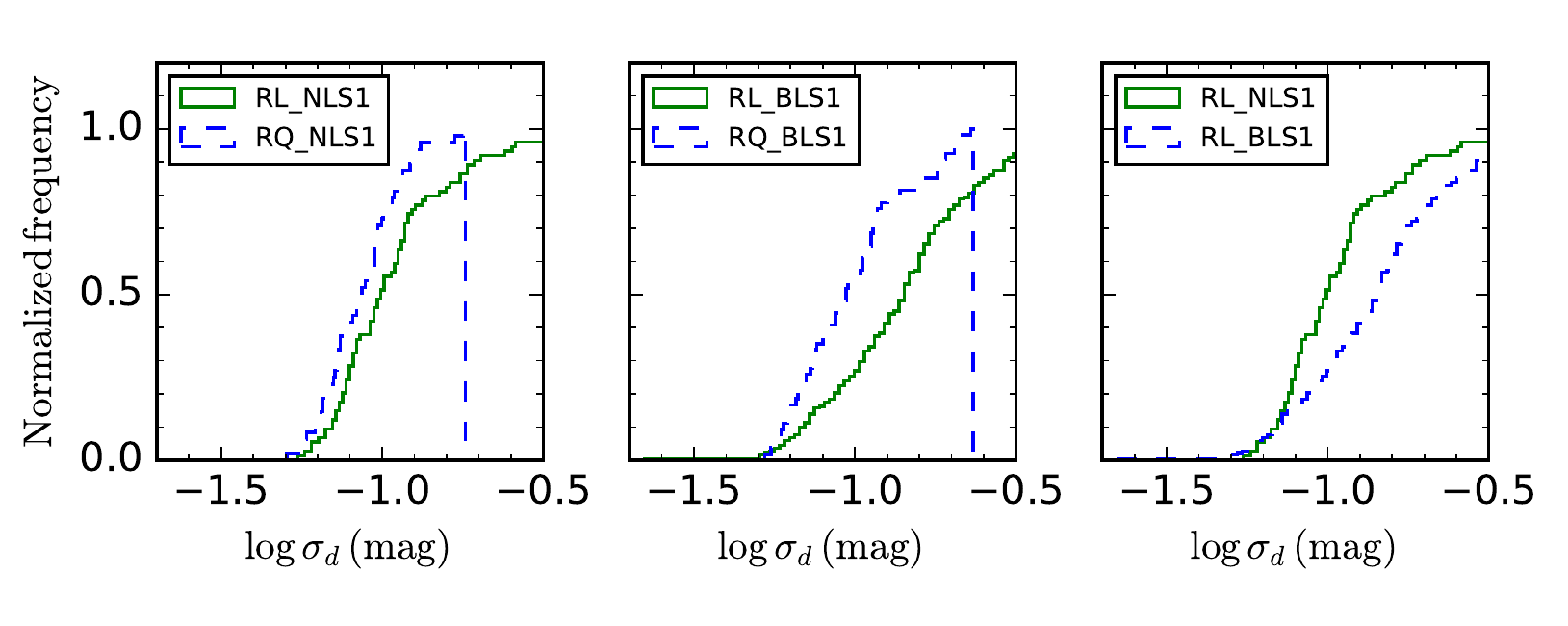}}
\caption{The cumulative distribution of radio-loud NLSy1 v/s radio-quiet NLSy1 galaxies (left), radio-loud BLSy1 v/s radio-quiet BLSy1 galaxies (middle), and radio-loud NLSy1 v/s radio-loud BLSy1 galaxies (right). The solid and dashed lines are for NLSy1 and BLSy1 galaxies respectively.}\label{Fig:Radio_NLS1_BLS1}. 
\end{figure}

The cumulative distributions of their $\sigma_d$ values are plotted in Figure 
\ref{Fig:Radio_NLS1_BLS1}. In both NLSy1 and BLSy1 galaxies (left and middle), radio-loud objects show more variability than their radio-quiet 
counterparts. Comparing the radio-loud sample of 
both NLSy1 and BLSy1 galaxies (right panel), we found 
RL-BLSy1 galaxies are more variable than RL-NLSy1 
galaxies. The median values of $\sigma_d$ distributions are 
$0.100^{+0.067}_{-0.025}$ ($0.144^{+0.107}_{-0.065}$) mag and 
$0.087^{+0.026}_{-0.021}$ ($0.094^{+0.060}_{-0.029}$) mag for RL-NLSy1 
(RL-BLSy1) and RQ-NLSy1 (RQ-BLSy1) galaxies respectively. 
When the median variability amplitudes are compared, we find that 
within error bars, both radio-loud and radio-quiet sources have 
similar variability amplitudes. However, K-S test indicates that their
intrinsic distributions are different. A two-sample 
K-S test applied on the $\sigma_d$ distributions of RL-NLSy1 and RQ-NLSy1, 
RL-BLSy1 and RQ-BLSy1, as well as RL-NLSy1 and RL-BLSy1 galaxies yield a 
$p$-value of 4 $\times$ 10$^{-2}$ ($D$-statistics $=0.25$, left panel), 
1 $\times$ 10$^{-6}$ ($D$-statistics $=0.39$, middle panel) and 6 $\times$ 
10$^{-7}$ ($D$-statistics $=0.36$, right panel) respectively, confirming 
that the distributions are different.               
Therefore the OV of radio-loud sources must be due to some other
mechanisms in addition to variations caused due to accretion disk
instabilities that operate in radio-quiet sources. 

\begin{figure}
\centering
\resizebox{7cm}{5.5cm}{\includegraphics{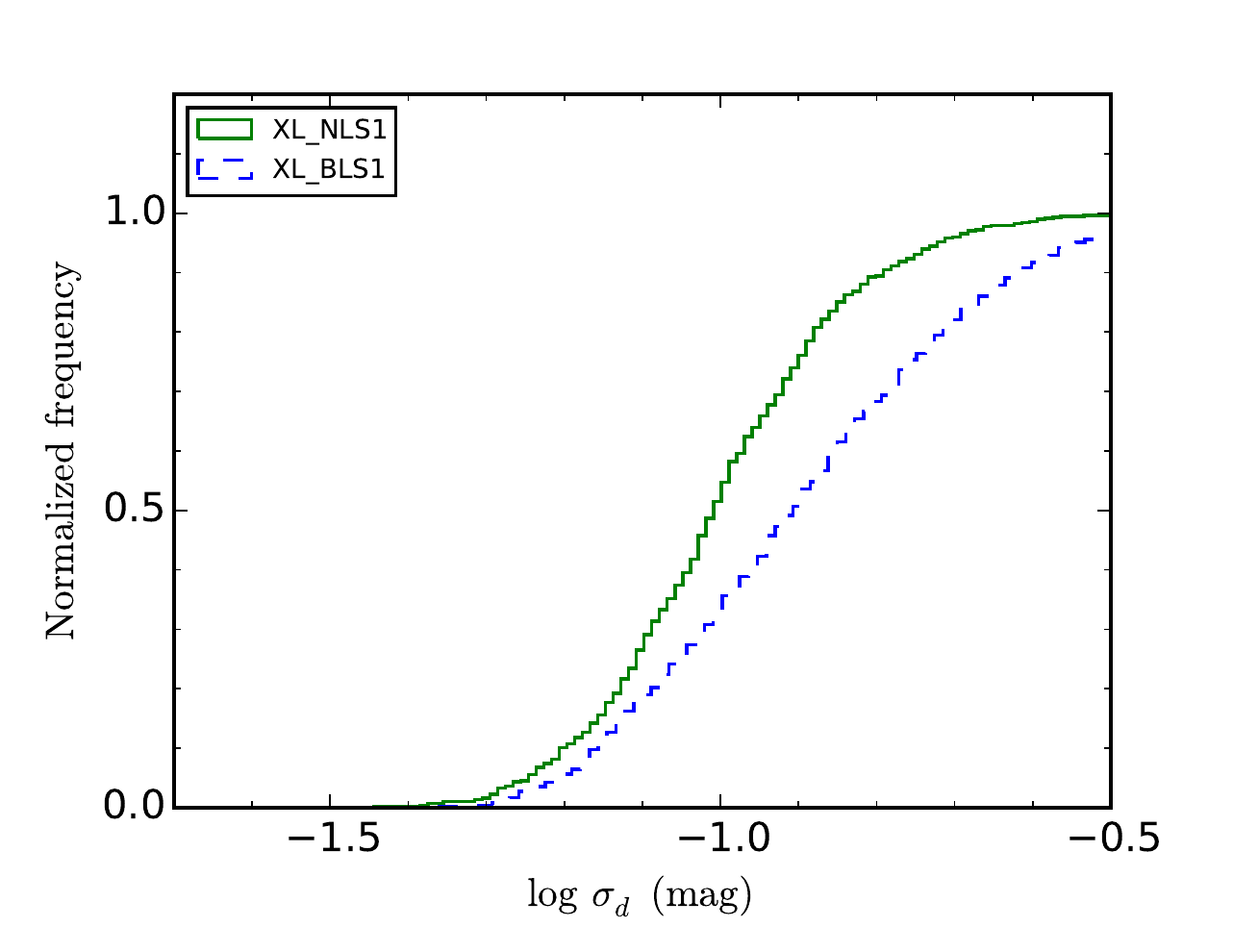}}
\caption{The cumulative distribution of X-ray detected BLSy1 (dashed line) and NLSy1 (solid line) galaxies.}\label{Fig:X-ray}. 
\end{figure}

 \begin{table*}
 \caption{Correlation of amplitude of variability ($\sigma_d$) with different AGN parameters. The columns are as follows: (1) variability parameters; (2) sample; (3) size of the sample. Columns (4)-(10) note the Spearman correlation coefficient (the $p$-value of no correlation) for the width of H$\beta$ line, $R_{5007}$, $R_{4570}$, $\log M_{\mathrm{BH}}/M_{\odot}$, $\log \lambda L_{5100}$, $\log \lambda_{\mathrm{Edd}}$ and redshift ($z$).}
	\begin{center}
 	\resizebox{\linewidth}{!}{%
     \begin{tabular}{ l l l r r r r r r r}\hline \hline 
      
    Test & Sample   & Size  &  FWHM(H$\beta$)  & $R_{5007}$ & $R_{4570}$ & $\log M_{\mathrm{BH}}/M_{\odot}$ & $\log \lambda L_{5100}$ &  $\log \lambda_{\mathrm{Edd}}$ & z \\ 
    (1)  & (2) & (3) & (4) & (5) & (6) & (7) & (8) & (9) & (10) \\ \hline
     $\sigma_d-$ &  NLSy1      & 2161	& $+$0.07(4e$-$04)  & $-$0.08(1e$-$04)  & $-$0.17(7e$-$15) & $+$0.13(1e$-$09) & $+$0.08(7e$-$05) & 0.00(9e$-$01) & $+$0.35(3e$-$65) \\
                 & BLSy1      & 2919 & $+$0.18(9e$-$25)    & $-$0.08(2e$-$06)  & $-$0.23(2e$-$37) & $+$0.27(3e$-$50) & $+$0.23(4e$-$38) & $-$0.07(2e$-$05) & $+$0.44(9e$-$137)\\
                 & All       & 5080 & $+$0.22(2e$-$60) & $-$0.09(3e$-$11) & $-$0.25(3e$-$73) & $+$0.27(3e$-$86) & $+$0.21(3e$-$55) & $-$0.16(1e$-$30) & $+$0.40(1e$-$200)\\
     \hline \hline
  
        \end{tabular} } 
        \label{Table:corr1}
        \end{center}
    \end{table*}
 
  \begin{figure*}
  \centering
  \resizebox{18cm}{7.0cm}{\includegraphics{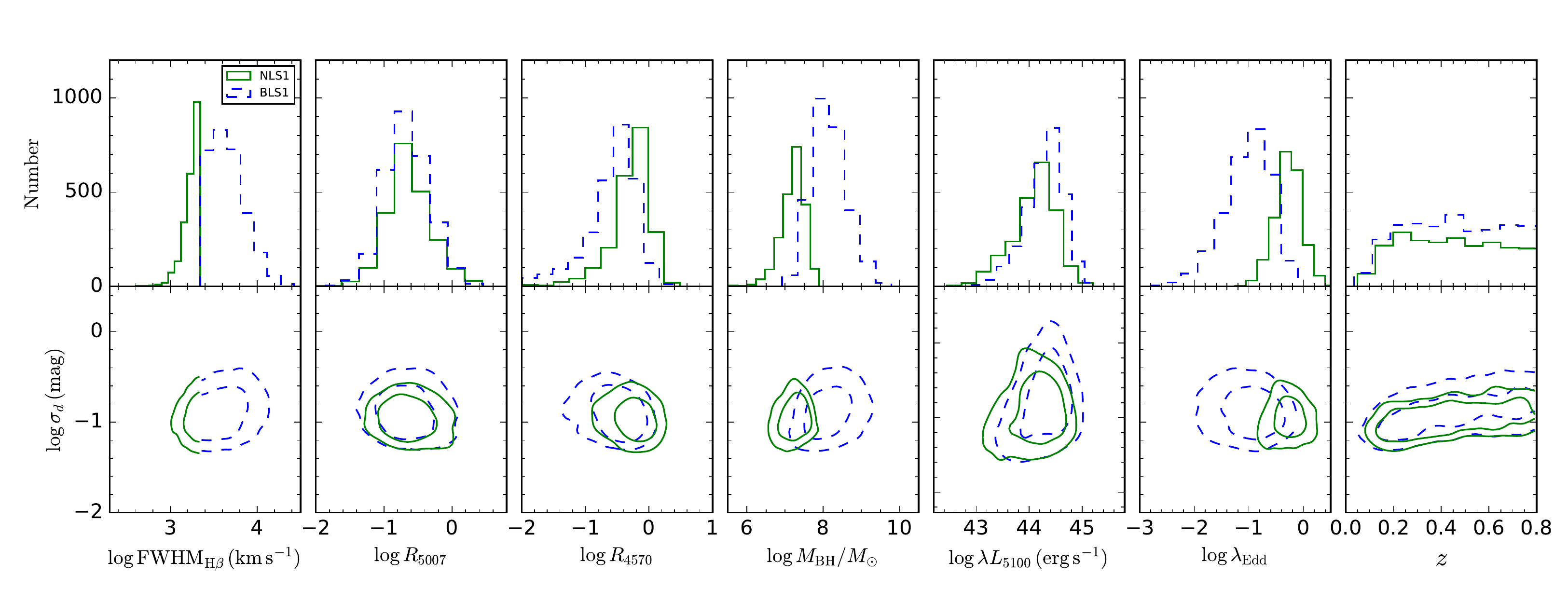}}
  \caption{From right to left, the correlations of $\sigma_d$ with FWHM$\mathrm{_{H\beta}}$, $R_{\mathrm{5007}}$, $R_{\mathrm{4570}}$, $M_{\mathrm{BH}}$, $\lambda L_{5100}$, $\lambda_{\mathrm{Edd}}$, and $z$ are plotted in the lower panels. The upper panels show the distribution of the parameters of all the objects in the sample i.e., the cut of the lower panels along the x-axis. The dashed contours represent BLSy1 galaxies while the solid contours correspond to NLSy1 galaxies. The shown contours are the 68 and 95 percentile density contours.}\label{Fig:NLS1_BLS1_prop}. 
  \end{figure*} 
  
\subsection{X-ray subsample}\label{sec:x-ray}
NLSy1 galaxies are known to show strong soft X-ray variability 
than their broad line counterparts \citep{1999ApJS..125..297L,2004AJ....127.1799G}. To ascertain if this 
nature of NLSy1 galaxies holds true in their OV properties too, we created a 
subsample of X-ray detected (XL) NLSy1 and BLSy1 galaxies from the work of \citet{2017ApJS..229...39R}. A total of 577 NLSy1 galaxies  and 653  BLSy1 galaxies, out of 2161 NLSy1 galaxies and 2919 BLSy1 galaxies are detected in the second ROSAT all-sky (2RXS) source catalog \citep{2016A&A...588A.103B}. The cumulative distribution of their $\sigma_d$ values is shown in 
Figure \ref{Fig:X-ray}. The $\sigma_d$ 
distribution has a median of $0.099^{+0.043}_{-0.027}$ mag and $0.124^{+0.084}_{-0.047}$ mag for XL-NLSy1 
and XL-BLSy1 galaxies respectively. A two-sample K-S test confirms distributions to be significantly different ($D$-statistics$=0.25$ and 
$p$-value$= 2\times$ 10$^{-18}$). Hence, the XL-BLSy1 galaxies are more variable than XL-NLSy1 galaxies. As the high flux variation shown by NLSy1 galaxies relative to BLSy1 galaxies in the X-ray band is not seen in the OV light curves, it is clear that the physical processes causing the X-ray flux variation and OV are different.  

\begin{center}
 \begin{table*}
 \caption{Correlation of amplitude of variability ($\sigma_d$) with different parameters as Table 1. The values given in columns 4-9 are the Spearman correlation coefficient (the $p$-value of no correlation).}
 	\resizebox{\linewidth}{!}{%
     \begin{tabular}{ l l r r r r r r r}\hline \hline   
   Sample  & redshift & Size  &  FWHM(H$\beta$)  & $R_{5007}$ & $R_{4570}$ & $\log M_{\mathrm{BH}}/M_{\odot}$ & $\log \lambda L_{5100}$ &  $\log \lambda_{\mathrm{Edd}}$ \\  \hline \hline
    NLSy1  & $0.0-0.2$ & 270 & $-$0.03(5e$-01$) & $-$0.11(5e$-$02) & $-$0.17(4e$-$03) & 0.04(5e$-$01) & 0.09(1e$-$01) & 0.07(2e$-$01) \\
    BLSy1  &           & 367 & 0.03(5e$-$01)    & $-$0.11(2e$-$02) & $-$0.17(1e$-$03) & 0.07(1e$-$01) & 0.14(4e$-$03) & 0.01(8e$-$01) \\
    All   &            & 637 & 0.01(8e$-$01)    & $-$0.11(2e$-$03) & $-$0.16(3e$-$05) & 0.04(3e$-$01) & 0.11(4e$-$03) & 0.01(6e$-$01) \\ \\
    
    NLSy1  & $0.2-0.4$ & 685  & 0.07(4e$-$02) & $-$0.03(4e$-$01) & $-$0.23(2e$-$09) & 0.12(7e$-$04) & 0.08(2e$-$02) & $-$0.01(8e$-$01)\\
    BLSy1  &           & 863  & 0.19(1e$-$08) & $-$0.08(1e$-$02) & $-$0.19(1e$-$08) & 0.21(1e$-$10) & 0.11(5e$-$04) & $-$0.16(2e$-$06) \\
    All    &           & 1548 & 0.21(4e$-$18) & $-$0.06(8e$-$03) & $-$0.25(1e$-$23) & 0.23(6e$-$21) & 0.13(1e$-$07) & $-$0.18(1e$-$13) \\ \\
    
    NLSy1 & $0.4-0.6$ & 612  & 0.09(1e$-$01) & 0.00(9e$-$01) & $-$0.24(1e$-$09) & $-$0.10(1e$-$02) & $-$0.30(1e$-$14) & $-$0.27(3e$-$12)\\
    BLSy1 &           & 847  & 0.20(3e$-$09) & 0.06(4e$-$02) & $-$0.39(3e$-$31) & 0.12(4e$-$04)    & $-$0.27(4e$-$16) & $-$0.25(2e$-$14) \\
    All   &           & 1459 & 0.28(2e$-$28) & 0.03(1e$-$01) & $-$0.38(2e$-$51) & 0.21(1e$-$16)    & $-$0.17(8e$-$11) & $-$0.34(6e$-$42) \\ \\ 
     
    NLSy1 & $0.6-0.8$ & 571&$-$0.00(9e$-$01) & 0.09(1e$-$02) & $-$0.29(8e$-$13) & $-$0.35(4e$-$18) & $-$0.49(2e$-$36) & $-$0.33(1e$-$16)\\
    BLSy1 &           & 841  & 0.23(2e$-$12) & 0.10(2e$-$03) & $-$0.34(1e$-$24) & 0.11(6e$-$03)    & $-$0.42(7e$-$39) & $-$0.32(1e$-$22)\\
    All &             & 1412 & 0.27(3e$-$25) & 0.09(6e$-$04) & $-$0.37(3e$-$48) & 0.16(2e$-$10)    & $-$0.32(4e$-$35) & $-$0.37(2e$-$49)\\ \hline \hline
       
    \end{tabular} }
     	\label{Table:corr2}
\end{table*}
\end{center}

\subsection{Correlation of variability and emission line parameters}
The work of \citet{2017ApJS..229...39R} has yielded various emission line parameters
of the sources in our sample. To understand how variability is related to the 
key physical properties of AGN, we tested several correlations between 
variability and various physical characteristics of the sources such as
FWHM(H$\beta$), strength of [O {\small III}] line (defined as
$R_{5007}=F_\mathrm{[O {\small III}]} (5007\AA)/\mathrm{H}
\beta_{\mathrm{tot}}$), Fe II strength relative to H$\beta$ (defined as 
$R_{4570}=$ Fe {\small II}($\lambda 4434-4684$)/H$\beta_b$),  
and monochromatic luminosity at 5100$\AA$ ($\lambda L_{5100}$). 
All these parameters were taken from the work of \citet{2017ApJS..229...39R}. In Figure 
\ref{Fig:NLS1_BLS1_prop}, the distributions of individual parameters are 
shown in the top panel, while in the bottom panel the correlations 
of them with $\sigma_d$ is shown. Table \ref{Table:corr1} 
summarizes the Spearman's rank correlation coefficient and the two-sided 
$p$-values for the null hypothesis of no correlation.

The amplitude of variability is positively correlated with the width of H$\beta$ line but negatively correlated  with $R_{4570}$. 
Spearman's rank correlation test confirms both the correlations to be significant. 
The above correlations are found to remain when all the sources
are considered together or separately for the sample of 
BLSy1 and NLSy1 galaxies.  The strong anticorrelation of variability amplitude with $R_{4570}$ found here implies a lower variability in NLSy1 than BLSy1 galaxies since the former has stronger Fe II emission than the latter. However, no correlation between $\sigma_d$ and $R_{5007}$ has been found. These results are consistent with the findings of \citet{2010ApJ...716L..31A} although their sample is very small compared 
to this work. It is likely that the low amplitude of variability in NLSy1
galaxies compared to BLSy1 galaxies is an outcome of the correlation seen
between $\sigma_d$ and width of H$\beta$ line and $R_{4570}$.
A positive correlation is also found between $\sigma_d$ and $\lambda L_{5100}$.

\subsection{Dependence of variability with redshift}
Though the redshift range of this study is limited to $z<0.8$ \citep[demanded 
by the presence of both H$\alpha$ and H$\beta$ in the 
SDSS spectra;][]{2017ApJS..229...39R}, a strong positive correlation is found between $\sigma_d$ and $z$ with a Spearman's rank correlation coefficient of 0.40 and $p$-value of 
1 $\times$ 10$^{-200}$ with the high $z$ sources showing larger amplitude of 
variability than their low $z$ counterparts. This is the strongest 
correlation among all the other correlations investigated here. Such redshift evolution of variability was also noticed in quasars
by \citet{2004ApJ...601..692V} up to $z \sim 5$ although 
\citet{2010ApJ...708..927K} and \citet{2010ApJ...721.1014M} found a negligible 
trend with redshift suggesting that the variability is intrinsic to the quasar 
and do not evolve over cosmic time for fixed physical parameters of the 
quasars (black hole mass, absolute magnitude etc). Since AGNs are more variable at shorter wavelengths and therefore, positive correlation observed between $\sigma_d$ and $z$ in this work is most likely a manifestation of the anticorrelation known between variability and wavelength \citep{1996MNRAS.282.1191C} because higher redshifts probe shorter rest frame wavelengths.

\subsection{Dependence of variability with $M_{\mathrm{BH}}$}
In the process of selection of new NLSy1 galaxies, \citet{2017ApJS..229...39R} has
carried out spectral fitting of SDSS spectra for all the candidates selected
in their study. The results of that fitting was used to derive the black hole
mass of each of the sources assuming virial relationship using the following
equation
\begin{equation}
M_{\mathrm{BH}} = f R_{\mathrm{BLR}} \Delta v^2/G
\end{equation}
where, $\Delta v$ is the FWHM of the broad component of
H$\beta$ emission line and $f$ is a scale factor that depends strongly on the geometry and kinematics of the BLR \citep{2015MNRAS.447.2420R}. Considering the spherical distribution of clouds we used $f = 3/4$. In Figure \ref{Fig:NLS1_BLS1_prop} the correlation between variability amplitude
and BH mass is shown. It is clear from the figure that $\sigma_d$ is positively
correlated with $M_{\mathrm{BH}}$.
The positive correlation found here between $\sigma_d-M_{\mathrm{BH}}$ 
is consistent with \citet{2007MNRAS.375..989W}, \citet{2008MNRAS.383.1232W} 
and \citet{2010ApJ...716L..31A}. 
However, \citet{2010ApJ...716L..31A} found the 
correlation to vanish when the dependency of $\mathrm{\lambda_{Edd}}$ is 
considered in the relation. \citet{2008MNRAS.387L..41L} suggested that such a positive 
correlation between variability amplitude and black hole mass can be explained 
in terms of an accretion disk model having the mean accretion rate of 
$\dot{m_o}=0.1$ and a variation of $0.1-0.5 \dot{m_o}$.

 \begin{figure}
 \centering
 \resizebox{9cm}{8.0cm}{\includegraphics{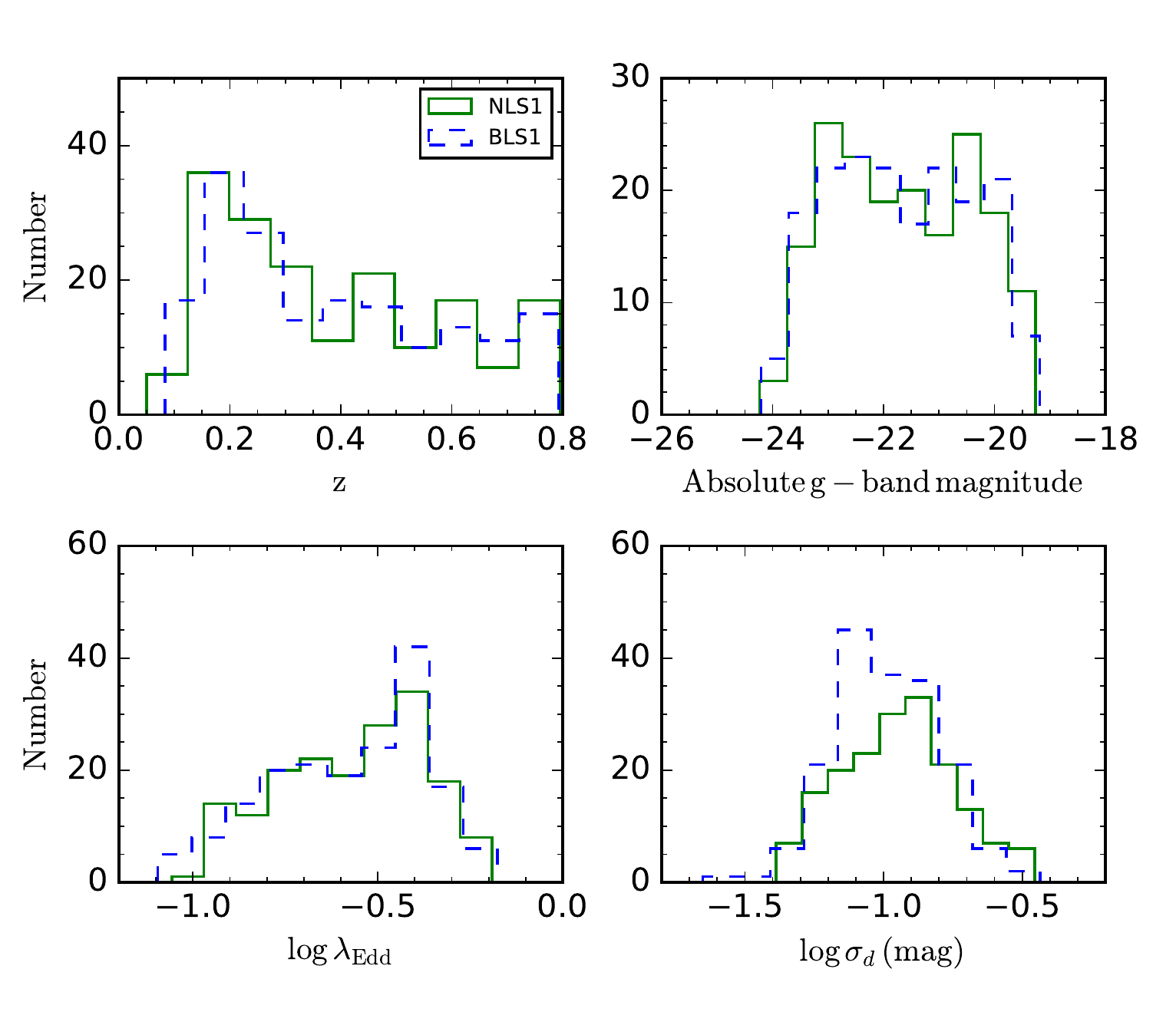}}
 \caption{The distribution of $z$ (upper-left), $M_g$ (upper-right), $\lambda_{\mathrm{Edd}}$ (lower-left) and $\sigma_d$ (lower-right) for a sub-sample of NLSy1 (solid) and BLSy1 (dashed) galaxies.}\label{Fig:var_edd}. 
 \end{figure}

\subsection{Dependence of variability with Eddington ratio}
The Eddington ratio, defined as the ratio of the bolometric
luminosity to Eddington luminosity  is a very important physical parameter
that characterizes the accretion rate of an AGN. For the sources in our sample,
Eddington ratio ($\lambda_{\mathrm{Edd}}$) is estimated as

\begin{equation}
\lambda_{\mathrm{Edd}} = L_{\mathrm{bol}}/L_{\mathrm{Edd}}
\end{equation}

where $L_{\mathrm{bol}}= 9 \times \lambda L_{\lambda} \mathrm{(5100\AA)} \, \mathrm{erg\, s^{-1}}$ and $L_{\mathrm{Edd}} = 1.3 \times 10^{38}\, M_{\mathrm{BH}}/M_{\odot} \, \mathrm{erg\, s^{-1}}$ \citep{2000ApJ...533..631K}. The correlation between $\sigma_d$ 
and $\lambda_{\mathrm{Edd}}$ is shown in Figure \ref{Fig:NLS1_BLS1_prop}, wherein
an anti-correlation is observed. This correlation will have effects due to
uncertainties in calculation of $M_{\mathrm{BH}}$ and consequently $L_{\mathrm{Edd}}$.
Such a correlation is also observed by \citet{2009ApJ...698..895K}, \citet{2010ApJ...721.1014M}, \citet{2010ApJ...716L..31A} and many others. It has been shown by \citet{2010ApJ...716L..31A} 
that the correlation between $\sigma_d$ and $\mathrm{\lambda_{Edd}}$ remains 
significant even after taking the dependency of $M_{\mathrm{BH}}$ leading
them to conclude the existence of a robust negative correlation 
between $\sigma_d$ and $\lambda_{\mathrm{Edd}}$.  

To study the effect of $\lambda_{\mathrm{Edd}}$ on $\sigma_d$, we created a subsample of 176 NLSy1 and BLSy1 galaxies each, by matching as close as possible their $z$, $M_g$ and $\lambda_{\mathrm{Edd}}$. The distribution of $z$ (upper-left), $M_g$ (upper-right) and $\lambda_{\mathrm{Edd}}$ (lower-left) for this sub-sample of NLSy1 (solid) and BLSy1 (dashed) galaxies are shown in Figure \ref{Fig:var_edd}. The distributions look similar. A K-S test gave $D$-statistics ($p$-values) of 0.04 (0.99), 0.05(0.93) and 0.05 (0.93) for the distributions of $z$, $M_g$ and $\lambda_{\mathrm{Edd}}$, that indicates of no differences in the distributions of $z$, $M_g$ and $\lambda_{\mathrm{Edd}}$ between the sub-samples of NLSy1 and BLSy1 galaxies. Also, the distribution of $\sigma_d$ (lower-right) for this sub-sample is shown in Figure \ref{Fig:var_edd}. A two sample K-S test applied to the distributions of $\sigma_d$ for this subsample of NLSy1 and BLSy1 galaxies gives a $D$-statistics and $p$-value of 0.10 and 0.30 respectively indicating that this sub-sample of NLSy1 and BLSy1 galaxies have similar $\sigma_d$ distributions. Therefore, when matched in $\lambda_{\mathrm{Edd}}$, both NLSy1 and BLSy1 galaxies are indistinguishable in their amplitude of OV. However, when considering the full sample of NLSy1 and BLSy1 galaxies, the larger amplitude of OV shown by BLSy1 galaxies relative to their NLSy1 counterparts is due to them having lower $\lambda_{\mathrm{Edd}}$ compared to NLSy1 galaxies which is also manifested in the negative correlation between $\sigma_d$ and $\lambda_{\mathrm{Edd}}$.

The correlation between $\sigma_d$ and $\mathrm{\lambda_{Edd}}$ found here
can be understood from the simple standard accretion disk model 
\citep{1973A&A....24..337S}. If the 
emission originates from the inner accretion disk, the emission decreases 
as it propagates outward. As the Eddington ratio increases, the radius ($r$) of 
emission region at a given wavelength moves outward i.e., $r$ increases with 
Eddington ratio since $r \sim T^{-1} \sim (\dot{m}/M_{\mathrm{BH}})^{1/3} \lambda^{4/3}$, where 
$T$ is the temperature of the disk, $\lambda$ is the wavelength and $\dot{m}$ is the mass accretion rate 
normalized by the Eddington rate. Since NLSy1 galaxies have higher Eddington ratio 
compared to BLSy1 galaxies at a given wavelength, the size of emission region 
is larger in NLSy1 galaxies than BLSy1 galaxies and thus variability amplitude is 
lower in the former than the latter.          

\begin{figure*}
\centering
\resizebox{18cm}{3.0cm}{\includegraphics{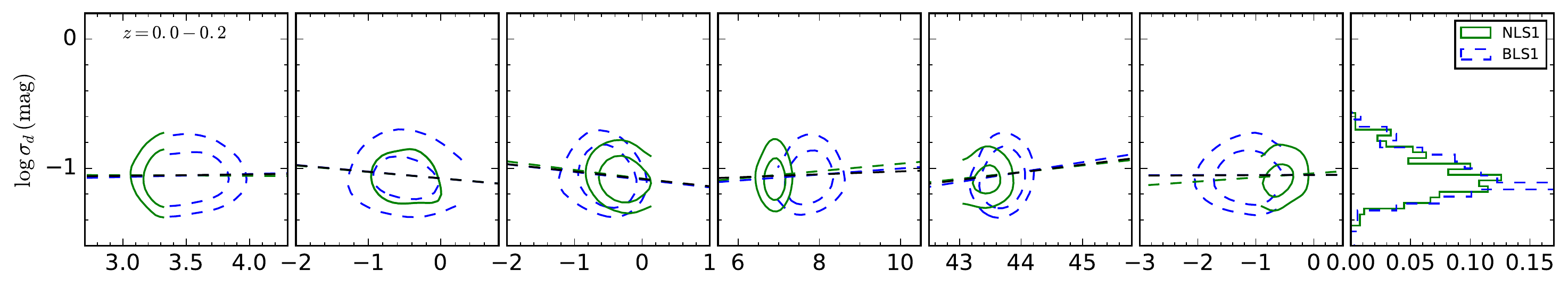}}
\resizebox{18cm}{3.0cm}{\includegraphics{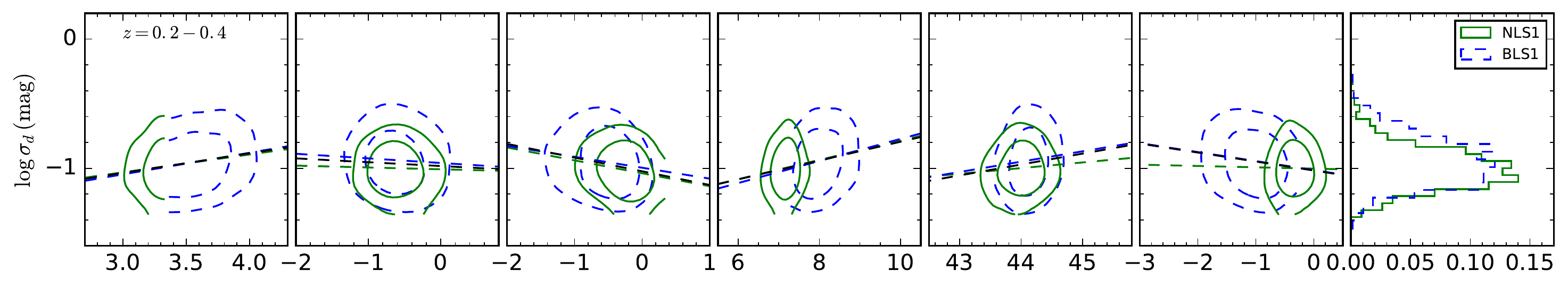}}
\resizebox{18cm}{3.0cm}{\includegraphics{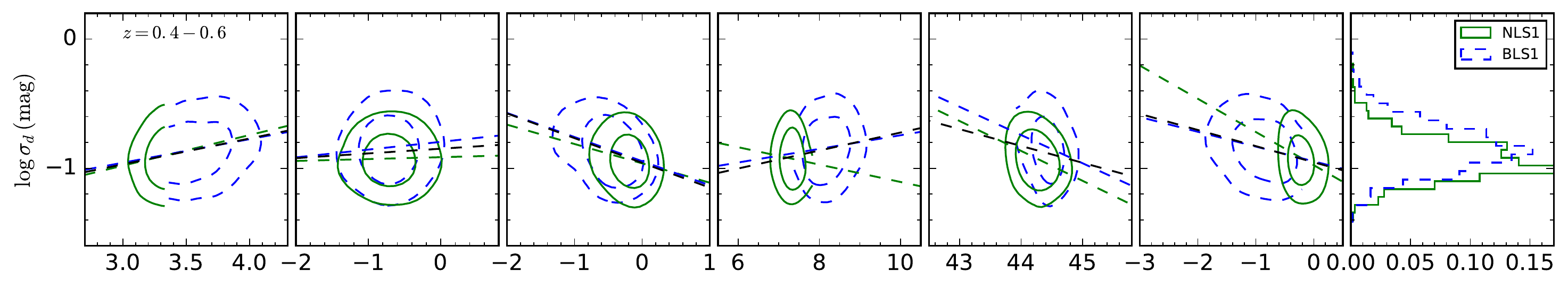}}
\resizebox{18cm}{3.3cm}{\includegraphics{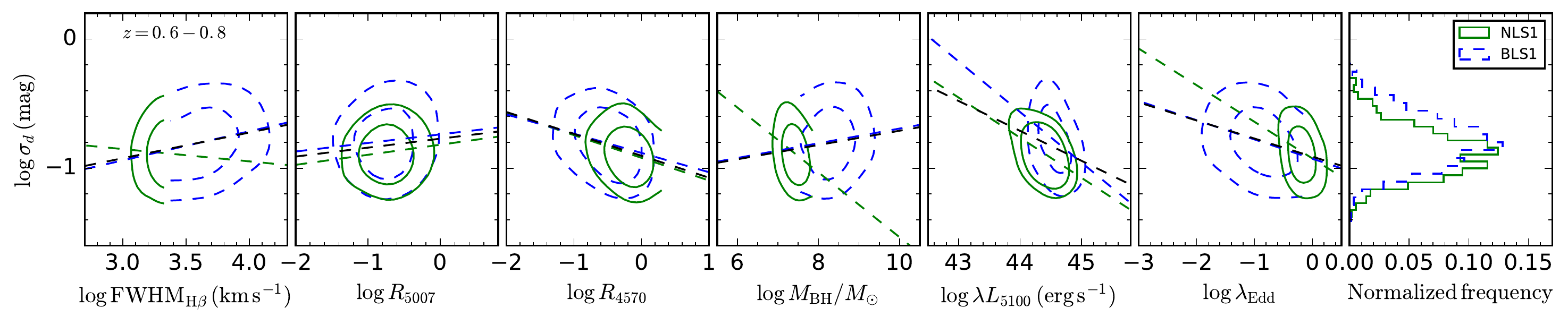}}
\caption{From right to left, the correlation of $\sigma_d$ with FWHM$\mathrm{_{H\beta}}$, $R_{\mathrm{5007}}$, $R_{\mathrm{4570}}$,  $M_{\mathrm{BH}}$, $\lambda L_{5100}$, $\lambda_{\mathrm{Edd}}$, and normalized distribution of $\mathrm{\sigma_d}$ is plotted for redshift $z=0.0-0.2$ (top), $z=0.2-0.4$ (upper-middle), $z=0.4-0.6$ (lower-middle), and $z=0.6-0.8$ (bottom). The dashed contours represent BLSy1 galaxies while the solid contours corresponds to NLSy1 galaxies. The shown contours are 68 and 95 percentile density contours.}\label{Fig:var_par_z}. 
\end{figure*}

\subsection{Variability vs physical parameters in redshift bins}
To study the correlations mentioned in the earlier sections in detail, we further divided the sample into different redshift bins 
($z=0.0-0.2$, $0.2-0.4$, $0.4-0.6$, $0.6-0.8$). All the correlations are shown in Figure \ref{Fig:var_par_z} and results of the correlation analysis are given in Table \ref{Table:corr2}. It 
seems that at lower redshifts ($z<0.4$), the correlation between $\sigma_d$ and all the physical parameters investigated here is insignificant but for 
higher redshifts ($z>0.4$), $\sigma_d$ is strongly correlated 
with FWHM(H$\beta$), $R_{4570}$, $\lambda L_{5100}$ 
and $\mathrm{\lambda_{Edd}}$. $\sigma_d$ increases 
with FWHM(H$\beta$) but decreases with Fe II strength, luminosity, and 
Eddington ratio. Analyzing the correlation between $\sigma_d$ and $M_{\mathrm{BH}}$ we found it to be weak when dividing all the objects in different redshift bins. 

\subsection{Correlation of variability and radio-loudness}
\begin{figure}
\centering
\resizebox{8cm}{4.0cm}{\includegraphics{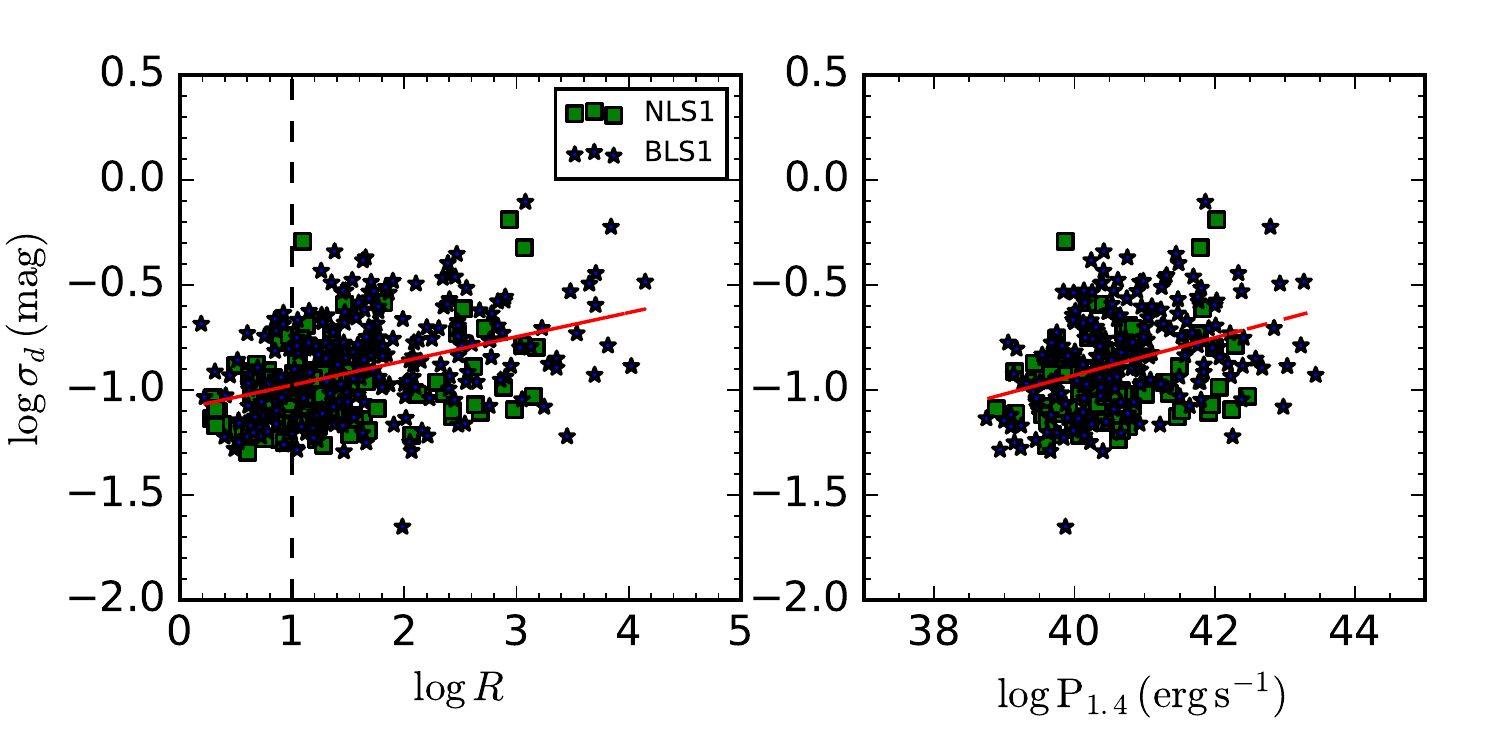}}
\caption{The $\sigma_d$ vs radio-loudness (left) and radio-power (right) is plotted. The dashed line at $\log\, R=1$ in the left panel is the dividing line between radio-loud and radio quiet objects. The solid lines represent the best linear fit to the relation. Spearman rank correlation coefficient ($p$-value) are 0.40 ($2\times 10^{-17}$) for $\sigma_d-R$ relation and 0.36 ($8\times 10^{-11}$) for $\sigma_d-P_{1.4}$ relation respectively considering total sample.}\label{Fig:Radio_loudness}. 
\end{figure}

A small sub-set of our sample of NLSy1 and BLSy1 galaxies have radio
counterparts from the FIRST survey. In both the population of BLSy1 and
NLSy1 galaxies radio-loud sources are found to be more variable than
radio-quiet sources (see section \ref{sec:radio}). Though Eddington ratio plays an important role an additional mechanism might be at work in radio-loud objects. Since the origin of radio emission is relativistic jets, some contribution of it might influence the optical variability. In Figure \ref{Fig:Radio_loudness}, we plotted the 
variability amplitude against radio-loudness (left panel) and radio-power (right panel) for both NLSy1 and BLSy1 galaxies. Interestingly,  a 
moderately strong correlation between variability amplitude and radio-loudness 
is found. This correlation is also present when the two 
samples are considered separately. The optical variability is also 
positively correlated with radio-power ($r=0.36$ and $p$-value=$8\times 10^{-11}$) suggesting that objects with strong jet show a large amplitude of variation in 
optical. The variation of $\sigma_d$ with radio-loudness and radio-power can be explained by the following relations

\begin{align}
\log \sigma_d &= &(0.11\pm 0.01) \log R + (-1.09 \pm 0.02) \\
			 &	= &(0.09 \pm 0.01) \log P_{1.4} +(-4.51\pm 0.56)
\end{align}
       
This finding leads us to hypothesize that the mechanisms
for OV in radio-loud and radio-quiet objects can be quite different. It is
likely that the optical emission in radio-quiet sources is due to
the presence of both non-thermal emission from the jet in addition to the
thermal emission from the accretion disk. Alternatively, in radio-quiet
sources, the optical emission is due to accretion disk thermal emission.

\section{Summary and Conclusion}\label{sec:conclusion}
We studied the OV of a large sample of NLSy1 and BLSy1 galaxies using archival $V$-band data from CRTS. The present study is a manifold increase compared to the earlier work in terms of (a) the number of objects used, (b) the epochs of data used for each of the objects, and (c) the duration of observations. In this work we have used a sample of 5510 NLSy1 and 5510 BLSy1 galaxies well matched in the $M_g-z$ plane. Each of these objects has a minimum of 50 epochs of data spanning 5 to 9 years. Therefore, the present sample along with the rich data set is ideal for a comparative study of the OV properties of NLSy1 and BLSy1 galaxies. The $V$-band light curves of our sample sources were modeled using DRW to understand their variability. From our sample, 2161 (39.2\%) NLSy1 and 2919 (52.9\%) 
BLSy1 show variability in CRTS long-term light curve on time scale larger than 
a day. The sources that showed variability on time scale larger than a day
are further considered for detailed analysis. Our main findings 
are as follows.
\begin{itemize}
\item The median amplitude of variability is 
found to be $0.107^{+0.057}_{-0.032}$ mag for NLSy1 galaxies and $0.129^{+0.082}_{-0.049}$ for BLSy1 galaxies. Though the median values of $\sigma_d$ agree
within error bars, a K-S test confirms with high significance that the two distributions are indeed different. Thus, NLSy1 galaxies as a class show lower amplitude of variation in 
the optical than BLSy1 galaxies. However, in a sub-sample of NLSy1 and BLSy1 galaxies that have nearly identical $\lambda_{\mathrm{Edd}}$, the distribution of $\sigma_d$ is found to be similar. Therefore, the larger amplitude of OV seen in BLSy1 galaxies relative to NLSy1 galaxies is due them having lower $\lambda_{\mathrm{Edd}}$ than NLSy1 galaxies.

\item Radio-loud objects in our sample in general are found to be more variable than their radio-quiet counterparts. Also, radio-loud BLSy1 galaxies are more variable 
than radio-loud NLSy1 galaxies as confirmed by K-S test. This increased variability in radio-loud sources both in NLSy1 and BLSy1 
galaxies relative to radio-quiet sources might be due to the presence
of non-thermal jet emission in addition to the thermal disk emission 
in them, compared to the contribution of only thermal emission 
from the accretion disk to 
the optical light in radio-quiet sources.

\item When X-ray detected NLSy1 and BLSy1 galaxies are considered separately, 
we find median amplitude of 
variations of $0.099^{+0.043}_{-0.027}$ mag for XL-NLSy1 and $0.124^{+0.084}_{-0.047}$ mag for XL-BLSy1 galaxies. According to K-S test XL-BLSy1 galaxies are more variable than XL-NLSy1 galaxies. 
 
\item A strong anti-correlation is found between the amplitude of variability 
and $R_{4570}$ and $\lambda_{\mathrm{EDD}}$ suggesting accretion disk as the main driver of the OV in both broad and narrow line Seyfert 1 galaxies.  

\item The amplitude of OV is found to be correlated with radio-loudness and 
radio-power. This hints for the contribution of jets in the OV of RL-NLSy1 and RL-BLSy1 galaxies in addition to Eddington ratio which is the main factor of OV in their radio-quiet counterparts.
\end{itemize}        

\acknowledgments
We are grateful for the comments and suggestions by the anonymous referee, which helped to improve the manuscript. We thank Ying Zu for useful discussions on the JAVELIN code. S.R. thanks Neha Sharma for carefully reading this manuscript. 

 \bibliographystyle{apj}
 \bibliography{ref}

\end{document}